\documentclass{aa} 

\usepackage{times,epsfig,amssymb,amsmath,natbib}

\usepackage{placeins}

 \usepackage{url}

\usepackage{float}
\usepackage[toc,page]{appendix}
\usepackage{tabularx}

\usepackage{natbib,twoopt}
\usepackage{verbatim}

\usepackage[]{hyperref}
\usepackage{hypcap}
\usepackage{txfonts}
\usepackage{color}
\usepackage{graphicx}	
\usepackage{amsmath}	
\usepackage{amssymb}	
\usepackage{xcolor}
 \usepackage{upgreek}

\begin{document}

\mail{mbrueggen@hs.uni-hamburg.de}

\title{Radio observations of the merging galaxy cluster system\\ Abell 3391-Abell 3395}
\titlerunning{Abell 3391/95}
\author{M. Br\"uggen\inst{1} 
\and
T.H. Reiprich\inst{2}
\and
E. Bulbul\inst{3}
\and
B. S. Koribalski\inst{4,9}
\and
H. Andernach\inst{5}
\and
L. Rudnick\inst{6} 
\and
D. N. Hoang\inst{1}
\and
A. G. Wilber\inst{7,1}
\and
S.~W. Duchesne\inst{7}
\and
A. Veronica\inst{2}
\and
F. Pacaud\inst{2}
\and
A.M. Hopkins\inst{8}
\and
R. P. Norris\inst{9,4}
\and
M. Johnston-Hollitt\inst{7,15}
\and 
M. J. I. Brown\inst{10}
\and
A. Bonafede\inst{13,14,1}
\and
G. Brunetti\inst{14}
\and
J. D. Collier\inst{11,12}
\and
J. S. Sanders\inst{3}
\and
E. Vardoulaki\inst{16}
\and
T. Venturi\inst{14}
\and
A. D. Kapinska\inst{17}
\and
J. Marvil\inst{17}}

\institute{
University of Hamburg, Hamburger Sternwarte, Gojenbergsweg 112, 21029 Hamburg, Germany
\and
Argelander-Institut f\"ur Astronomie, Universit\"at Bonn, Auf dem H\"ugel 71, Germany
\and
Max-Planck-Institut f\"ur extraterrestrische Physik, Giessenbachstra{\ss}e 1, 85748 Garching, Germany
\and
CSIRO Astronomy \& Space Science, P.O. Box 76, Epping, NSW 1710, Australia
\and
Depto.\ de Astronom{\'{i}}a, Universidad de Guanajuato, Callej\'on de
Jalisco s/n, Guanajuato, C.P.\ 36023, GTO, Mexico
\and
Minnesota Institute for Astrophysics, University of Minnesota, 116 Church St. SE, Minneapolis, MN 55455, USA
\and
International Centre for Radio Astronomy Research (ICRAR), Curtin University, Bentley, WA 6102, Australia
\and 
Australian Astronomical Optics, Macquarie University, 105 Delhi Rd, North Ryde, NSW 2113, Australia 
\and
Western Sydney University, Locked Bag 1797, Penrith, NSW 2751, Australia
\and 
School of Physics \& Astronomy, Monash University, Clayton, VIC 3800, Australia
\and
The Inter-University Institute for Data Intensive Astronomy (IDIA), Department of Astronomy, University of Cape Town, Private Bag X3, Rondebosch, 7701, South Africa
\and
School of Science, Western Sydney University, Locked Bag 1797, Penrith, NSW 2751, Australia
\and
DIFA - Universit\'a di Bologna, via Gobetti 93/2, I-40129 Bologna, Italy
\and
INAF, Istituto di Radioastronomia, Via Gobetti 101, 40129 Bologna, Italy
\and
Curtin Institute for Computation, Curtin University, GPO Box U 1987, Perth, WA 6845, Australia
\and
Th\"{u}ringer Landessternwarte, Sternwarte 5, 07778 Tautenburg, Germany
\and
NRAO, PO Box 0, Socorro 87801, NM, USA
}

\abstract
{The pre-merging system of galaxy clusters Abell 3391-Abell 3395 located at a mean redshift of 0.053 has been observed at 1~GHz in an ASKAP/EMU Early Science observation as well as in X-rays with eROSITA. The projected separation of the X-ray peaks of the two clusters is $\sim$50$'$ or $\sim$ 3.1 Mpc.
Here we present an inventory of interesting radio sources in this field around this cluster merger. 
While the eROSITA observations provide clear indications of a bridge of thermal gas between the clusters, neither ASKAP nor MWA observations show any diffuse radio emission coinciding with the X-ray bridge. 
We derive an upper limit on the radio emissivity in the bridge region of $\langle J \rangle_{1\,{\rm GHz}}< 1.2 \times 10^{-44} {\rm W}\, {\rm Hz}^{-1} {\rm m}^{-3}$.
A non-detection of diffuse radio emission in the X-ray bridge between these two clusters has implications for particle-acceleration mechanisms in cosmological large-scale structure. We also report extended or otherwise noteworthy radio sources in the 30 deg$^2$ field around Abell 3391-Abell 3395. We identified 20 Giant Radio Galaxies, plus 7 candidates, with linear projected sizes greater than 1 Mpc. The sky density of field radio galaxies with largest linear sizes of $>0.7$ Mpc is $\approx 1.7$ deg$^{-2}$, three times higher than previously reported. We find no evidence for a cosmological evolution of the population of Giant Radio Galaxies. Moreover, we find seven candidates for cluster radio relics and radio halos.}
\maketitle


\section{Introduction}

Considerable efforts are being made to detect the Warm-Hot Intergalactic Medium (WHIM) in various bands of the electromagnetic spectrum. Detecting radio synchrotron emission from intercluster filaments or bridges could provide a new probe of the WHIM and could also shed light on mechanisms of particle acceleration and magnetic fields in a poorly studied environment \citep{2019A&A...627A...5V}. 
The magnetic field in the WHIM is less likely to be affected by a small-scale dynamo or by outflows from galaxies and active galactic nuclei (AGN). This field could therefore be a tracer of magnetic fields from an earlier epoch in the Universe that has subsequently been merely compressed by structure formation processes \citep[e.g.][]{2005ApJ...631L..21B, 2017CQGra..34w4001V}.

Moreover, interacting galaxy clusters produce diffuse radio sources by accelerating electrons via shocks or turbulence \citep[e.g.][]{2014IJMPD..2330007B}. Giant radio halos are diffuse radio synchrotron sources found in galaxy clusters that have recently suffered a merger, as indicated by a disturbed intracluster medium (ICM) or other indicators of the cluster’s dynamical state \citep[e.g.][]{2001A&A...378..408S, 2010A&A...517A..10C}. Cluster radio shocks or radio relics are arc-shaped, diffuse sources that appear to be related to shock waves in the ICM \citep{2020MNRAS.493.2306B}. Moreover, there is a class of sources that trace old radio plasma from AGN that has been re-energised through processes in the ICM. Low-frequency radio observations are starting to show more of these types of sources. The common properties of these sources are the AGN origin of the plasma and their ultra-steep radio spectra. Radio phoenices and gently reenergised tails (GReETs) are examples of such sources (see e.g. \cite{2019SSRv..215...16V} for a recent review).\\

The cluster Abell 3395 (hereafter A3395) is double-peaked, both in its galaxy distribution \citep{1997ApJ...482...41G} as well as in its X-ray emission, first observed with the Einstein satellite \citep{1981ApJ...243L.133F}. It is accompanied to the north by A3391 at a separation of about 3 Mpc. The mean redshift of A3395 ($z=0.0518$) and A3391 ($z=0.0555$) is $\langle z\rangle =0.053$. In between the clusters lies a galaxy group called ESO 161-IG 006 ($z=0.0520$, Alvarez et al. 2018). Both clusters have masses of around $M_{200}\sim 2\times 10^{14} M_\odot$ \citep{2011A&A...534A.109P}, and X-ray temperatures of around $kT\sim5$\,keV \citep{2009ApJ...692.1033V}. ASCA, ROSAT, Planck, and Suzaku observations have confirmed that A3395 and A3391 are connected by a gas bridge \citep{2001ApJ...563..673T, 2013A&A...550A.134P, 2017PASJ...69...93S}.


Diffuse radio emission from pairs of galaxy clusters is rare \citep{2018MNRAS.478..885B, 2019A&A...630A..77B, 2019Sci...364..981G, 2020MNRAS.tmpL.159B}. 
The first known bridge of low surface brightness connecting a radio halo and radio relic was found in the Coma cluster, first reported by \citet{1989Natur.341..720K} and further studied afterwards (e.g. \citealt{1990AJ.....99.1381V}, \citealt{2011MNRAS.412....2B}). A moderate correlation between radio and X-ray brightness has also been found in the Coma bridge. This correlation suggests that the radio and X-ray emission originate from the same volume and that the correlation is not produced by projection effects.

The bridges detected so far have different sizes and are observed in systems that are believed to be in a state preceding the merger. Currently, the best examples are the merging galaxy clusters Abell 399 and Abell 401, which are separated by about 5 Mpc.  Diffuse synchrotron emission was detected in the region connecting the two clusters using the Low Frequency Array (LOFAR) at $140$
MHz \citep{2019Sci...364..981G}. X-ray observations of A399-A401 have revealed a hot (6 - 7 keV)  filament of plasma in the region between the two clusters \citep[e.g.][]{2008PASJ...60S.343F}. Their masses are $5.7\times 10^{14}M_\odot$ and $9.3\times 10^{14}M_\odot$, respectively, and so the system is more than three times as massive as the A3391/5 system. The redshift of Abell 399 is 0.0718, while that of Abell 401 is 0.0736.
Here, both clusters host a radio halo. The presence of a bridge was
confirmed through the Sunyaev-Zeldovich effect \citep{2013A&A...550A.134P} but it is not seen in the GaLactic and Extragalactic All-Sky MWA (GLEAM) low-frequency radio survey \citep{ 2015PASA...32...25W}.

Integrating the average surface brightness over the Abell 399/401 bridge region, namely $\langle I \rangle_{140\,{\rm MHz}} = 2.75 \pm 0.08$ mJy beam$^{-1}$ (or 0.38 $\upmu$Jy arcsec$^{-2}$), over an area of $3 \times 1.3$ Mpc$^2$ and excluding the emission from the two halos, \cite{2019Sci...364..981G} obtain a total flux density of $S_{140\,{\rm MHz}} = 822 \pm 24$ mJy.
This flux density corresponds to a radio power of $L_{140\,{\rm MHz}} = 1.0 \times 10^{25}$ W Hz$^{-1}$ and a mean radio emissivity of $\langle J \rangle_{140\,{\rm MHz}} = 8.6 \times 10^{-44}$ W Hz$^{-1}$ m$^{-3}$. 

\cite{2020MNRAS.tmpL.159B} discovered a radio bridge in the pre-merging galaxy clusters A1758N and A1758S, which are 2 Mpc apart. The bridge is clearly detected in the LOFAR image at 144 MHz and tentatively detected at 53 MHz.
The clusters A1758N and A1758S also host radio halos but the mean radio emissivity in the bridge is more than one order of magnitude lower than that of the halos. Interestingly, radio and X-ray surface brightness 
are found to be correlated in the bridge.

The short lifetime for relativistic electrons emitting at 140 MHz limits the maximum distance that these relativistic electrons can travel in their lifetime to $<$0.1 Mpc \citep{2019Sci...364..981G}. Hence, the sizes of radio bridges combined with the short cooling time of the synchrotron-emitting electrons suggest that relativistic particles are accelerated in situ.  The large area filling factor, that is, a very smooth radio surface brightness, observed in the radio bridge of A399-A401 clearly disfavours shock acceleration as the main source of the observed radio emission.

The origin of radio synchrotron emission in radio bridges is not understood. Using cosmological simulations, \cite{2019Sci...364..981G} proposed that the synchrotron emission could come from electrons re-accelerated by weak shock waves provided there exists a volume-filling population
of pre-accelerated electrons in the bridge. However, this would not produce a radio-X correlation because shocks cover at most 10 percent of the volume of the bridge. Also, \cite{2019Sci...364..981G} concluded that the steep radio spectrum  challenges this scenario for the origin of the radio synchrotron bridge.

Recently, \citet{2020PhRvL.124e1101B} proposed that stochastic acceleration of relativistic electrons by turbulence could explain the radio emission observed in the system A399-A401. According to this model, the emission is expected to be volume-filling and with a steep spectrum\footnote{We define the spectral index $\alpha$ as $S \propto \nu^\alpha$.} ($\alpha <$ -1.3). Their model also predicts a natural correlation between radio and X-ray surface brightness. Moreover, it is expected that the radio emission gets clumpier at higher frequencies.

\begin{figure*}
\center{\includegraphics[width=1.9\columnwidth]{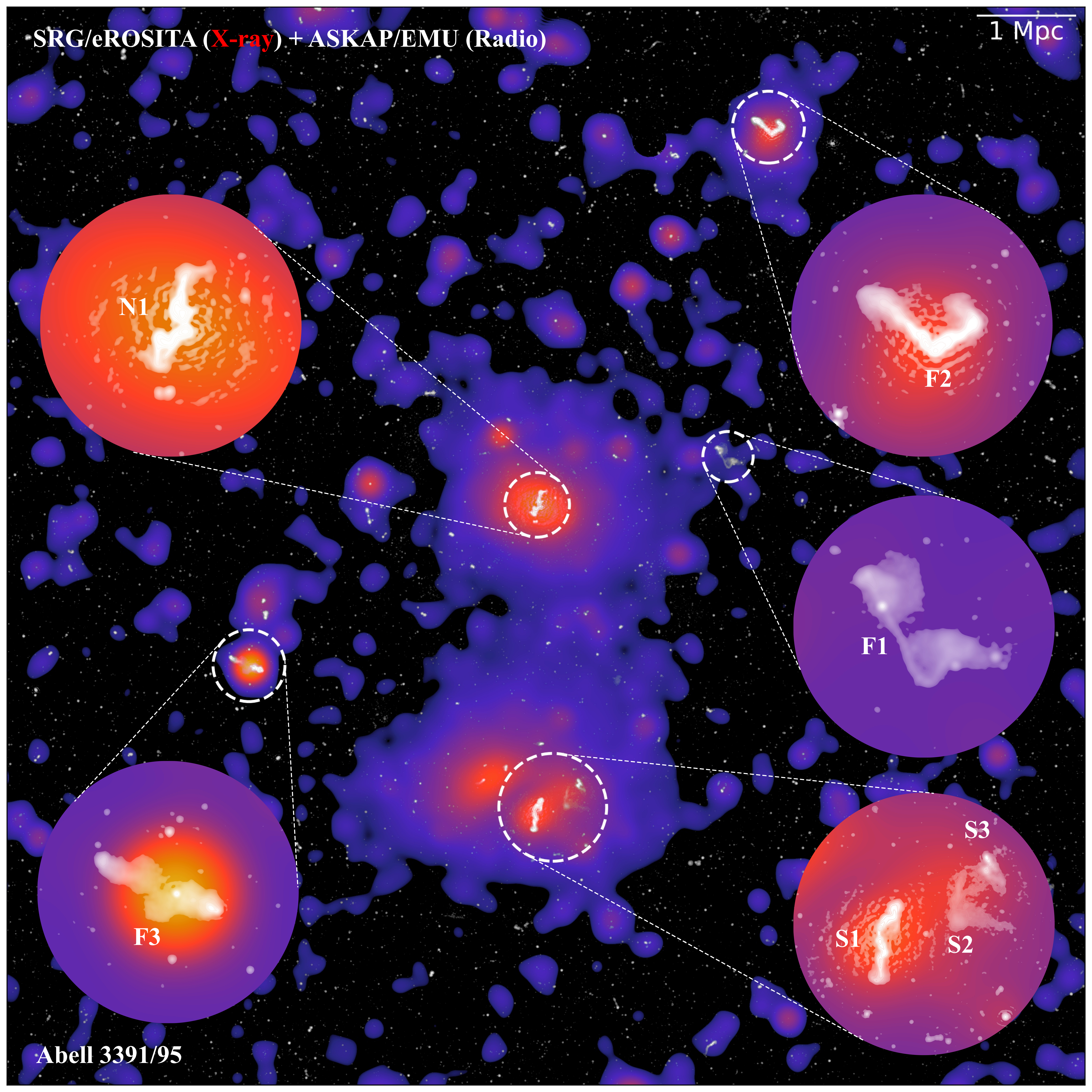}}
\caption{ASKAP/EMU radio and eROSITA X-ray overlay of the A3391-A3395 system. The eROSITA adaptively smoothed image covers the energy range from
0.3 to 2.0 keV. The inlays show notable individual radio galaxies, some of which show signs of interactions with the ambient medium. All sources highlighted and magnified in this image fall within the mean redshift of the cluster system except for F3 which is at $z=0.193$. Details of these sources are given in Tables~\ref{tab:RGs_EMU} and \ref{tab:flux}.}
\label{fig:overlay}
\end{figure*}

In this paper, we present the results of our search for diffuse radio synchrotron emission in the bridge of thermal gas between A3395 and A3391 that was recently observed with eROSITA. Moreover, we  present an inventory of interesting radio sources in this field around this cluster merger. The outline of this paper is as follows: In Sect.~\ref{sec:xray} we give a short overview of the eROSITA observations that are published in an accompanying paper (Reiprich et al. 2020). In  Sect.~\ref{sec:radio} we describe ASKAP and MWA observations. Results are presented in Sect.~\ref{sec:results} and we present our conclusions in Sect.~\ref{sec:conclusions}.

Throughout this paper, we use the fiducial cosmology $H_0 = 70$ km s$^{-1}$ Mpc$^{-1}$, $\Omega_M = 0.3$, and $\Omega_\Lambda = 0.7$. At the mean redshift of A3395 and A3391, 1 arcsec on the sky corresponds to $\sim 1.04$ kpc.

\section{X-ray observations}
\label{sec:xray}
On July 13, 2019, {eROSITA} (extended ROentgen Survey with an Imaging Telescope Array) was successfully launched and made its way into an L2 halo orbit \citep{Predehl2020}. During its Performance Verification (PV) phase, {eROSITA} surveyed the A3391-A3395 system as part of its first light observations. Reiprich et al. (2020) present the first results from four eROSITA PV phase observations of A3395-A3391. These cover a total area of about 15 deg$^2$, or about half the size of the accompanying Australian Square Kilometre Array Pathfinder \citep[ASKAP:][]{askap} field, with at least 30 s exposure ($\sim$10 deg$^2$ with at least 1000 s). The resulting X-ray image is shallower than the XMM-Newton observation of the A3395-A3391 system but covers a much wider area.

Also, owing to its better response to soft X-rays, in between A3395 and A3391, Reiprich et al. (2020) observe warm gas in emission in the bridge, which may constitute a WHIM detection. Emission along this bridge includes the galaxy group ESO 161-IG006, which produces only a small fraction of the total emission. While most gas in the bridge appears hot, these latter authors discover hints for cool primordial gas between the clusters. Moreover, several clumps of matter around the merging system were discovered. This is enabled by the X-ray hot gas morphologies and radio jet and lobe structures of their central AGN. Finally, Reiprich et al. (2020) discover a filamentary emission region north of and well beyond the virial radius of A3391 connecting to an apparently infalling cluster and a connected emission filament towards the south of A3395, suggesting a total length of the filament of $\sim$15 Mpc ($\sim$4 degrees). The Planck SZ Y-maps, as well as galaxy density maps from DECam observations \citep[DECam; see, e.g.][]{Flaugher_2015}, appear to confirm this new filament. For more information we refer the reader to Reiprich et al. (2020).

\section{Radio observations}
\label{sec:radio}
\subsection{ASKAP observations}

The A3391-A3395 system was observed as part of the Evolutionary Map of the Universe \citep[EMU:][]{emu} survey Early Science observations which use the new ASKAP telescope in Australia to make a census of radio sources in the sky south of +30$^\circ$ declination; see Fig.~\ref{fig:overlay} for an X-ray radio overlay. Our ASKAP observations reach a depth of 25--35 $\mu$Jy/beam root mean square (rms) at a spatial resolution of $\sim$10 arcsec in the frequency range 846.5--1134.5 MHz. These latter are therefore 15 times deeper than the NRAO VLA Sky Survey \citep[NVSS:][]{NVSS}, and have a five times better angular resolution (we note that the A3391-A3395 field is not covered by NVSS).

The ASKAP early science data covering A3395-A3391 are from Scheduling Block (SB) 8275 observed on March 22, 2019, and are publicly available at the \href{https://data.csiro.au/collections/#domain/casdaObservation/search/}{CSIRO ASKAP Science Data Archive} (CASDA; \citealp{2017ASPC..512...73C}). SB8275 contains 36 separate measurement sets corresponding to 36 separate beams, each with a unique phase-tracking centre. Each beam covers $\sim$1 deg$^{2}$ on the sky and all beams were observed simultaneously, giving an instantaneous field of view (FOV; after accounting for beam overlap) of $\sim$30 deg$^{2}$. Of the 36 ASKAP antennas, 35 were used for the observations, with a minimum baseline of $\sim$ 22 m and maximum baseline of $\sim$ 6.4 km. Each measurement set covers a  bandwidth of  288~MHz with a central frequency of 990.5~MHz. These data were processed through the official ASKAP processing pipeline (ASKAPsoft; \citealp{2019ascl.soft12003G}). The steps implemented in the ASKAPsoft pipeline will be described in an upcoming system description paper, but we refer the reader to the \href{https://www.atnf.csiro.au/computing/software/askapsoft/sdp/docs/current/pipelines/introduction.html}{CSIRO ASKAPsoft documentation} for further details. 

ASKAP forms beams electronically through a chequerboard Phased Array Feed at the prime focus of each antenna. Individual beams are then correlated between antennas. The ASKAP correlator generates 15552 spectral-line channels, which are then averaged to 288 1-MHz channels. 

Beam measurement sets are bandpass calibrated using a standard calibrator (PKS B1934--638)\footnote{Reynolds J. E., 1994, Technical report, A Revised Flux Scale for the AT Compact Array. ATNF, Epping}. The calibration observations occurred in a preceding scheduling block wherein PKS B1934--638 was placed at the centre of each beam for 200 seconds. The derived bandpass solutions also account for the other fundamental interferometric calibration terms, such as the flux-density, delay, and phase-referencing solutions. One cycle of frequency-independent, direction-independent, phase-only self-calibration is also applied for each beam as part of the imaging pipeline. 

Individual calibrated measurement sets are imaged and deconvolved separately with imaging weights determined by Wiener pre-conditioning. These weights have a setting equivalent to a Brigg’s robustness value of 0.0, which was determined to provide an optimal combination of resolution, sensitivity, and point spread function (PSF) quality. The $w$-projection technique is used to account for the non-coplanarity of the sky, using a total of 557 w-planes.  Multi-term deconvolution (with two terms) is used to model variation in source flux  over the synthesised bandwidth and multi-scale deconvolution (with six scales) is used to improve the modelling of extended sources. Beam images are then averaged with the same weight to form a mosaic in the image plane. Primary beam correction is applied during the linear mosaic process using circular Gaussian models whose size is determined by holography observations. Although the beams differ in shape and are coma-distorted at the edge of the FOV, an average value is used for all 36 beams with an estimated error of 10\%.

As part of the routine ASKAP/EMU validation procedures, a preliminary source catalogue was created from the mosaic image and compared with other large-area radio survey catalogues. For these data, compact radio sources were crossmatched with the Sydney University Molonglo Sky Survey \citep[SUMSS:][]{SUMSS} catalogue \citep{2003MNRAS.342.1117M}, resulting in 111 matches. The median astrometric offset of $\sim 1.3$ arcsec between these matches 
is consistent with zero offset given the dispersion in this measurement. The median flux ratio of the ASKAP sources versus their SUMSS counterparts is 0.81 overall but is approximately 1.0 for the brightest matches.  Given that this flux comparison is complicated by the large difference in spatial resolution and a modest difference in frequency (for which we assume a spectral index of -0.8) we do not interpret this result as evidence for a systematic flux error. However, there is one additional concern that contributes to the flux uncertainty in these data and that is the lack of a common restoring beam among the images.  Because the individual beam images are processed independently, the beams can have different UV coverage (e.g. because of flagging or declination) and therefore different PSFs. Gaussian fits to each of the individual 36 PSFs yielded beam sizes with geometric means (i.e. $\sqrt{b_{\rm maj}\,b_{\rm min}}$) ranging from 9.8 to 11.1 arcsec. As we were unable to propagate these individual beam sizes to the final linear mosaic image, we assumed a constant beam size of 11.2 $\times$ 9.5 arcsec at a position angle of 87 degrees (a geometric mean of 10.3 arcsec).  The difference between the true and assumed PSF size will lead to a systematic error in integrated flux measurements, which is typically only $\sim$3\% but could be as large as 10\% in some places in the mosaic.

The resulting images of the beams covering A3395-A3391, as well as the full ASKAPsoft mosaic image of SB8275, show artefacts that mainly emanate from two bright and extended radio galaxies at the centres of the clusters: sources N1 and S1 (see Fig.~\ref{fig:overlay} and \ref{fig:ASKAPradiomap}). Negative bowl-like artefacts are present around the radio galaxies, with larger-scale positive and negative rings extending up to $\sim 1$~deg from their centres. The cause of these artefacts is currently being investigated and will be addressed in an upcoming technical paper (Wilber et al. in prep). 

In an attempt to improve the image quality, \citet{2020arXiv200601833W}  recently tested direction-dependent calibration and imaging on ASKAP early science and pilot survey observations. These latter authors report a significant reduction in artefacts around bright, compact sources after implementing further data processing via the third-generation software packages KillMS (kMS) \citep{2014A&A...566A.127T, 2015MNRAS.449.2668S} and DDFacet (DDF) \citep{2018A&A...611A..87T}, which are officially being used for data products of the LOFAR Two-Metre Sky Survey (LoTSS; \citealp{2017A&A...598A.104S}). As part of their testing, \citet{2020arXiv200601833W} developed a direction-dependent ASKAP pipeline, which was used to create a mosaic image of SB8275. This direction-dependent mosaic image is referred to as the ASKAP-DD image hereafter and is used to measure flux densities of radio sources. 
We refer the reader to \citet{2020arXiv200601833W} and an upcoming paper, Wilber et al. in prep, for details of the ASKAP direction-dependent pipeline and specific processing steps. 

Finally, the convention for naming sources discovered with EMU is EMU NA JHHMMSS.s+DDMMSS where `N' will be replaced with `E' for `early science' or `P' for `pilot survey', and `A' is `S' for `source',  `C' for component, or `D' for `diffuse emission'.

\subsection{Murchison Widefield Array observations} 

\begin{table}
    \centering
    \caption{\label{tab:obs:mwa}Details of MWA-2 images used.}
    \begin{tabular}{c c c c c}
    \hline
    Band & $\nu_\text{c}$ & $\sigma_\text{rms}$ & $\Delta S_\nu$ & Beam \\\hline
    & (MHz) & (mJy\,beam$^{-1}$) & \% & (${}^{\prime\prime}\times{}^{\prime\prime}$) \\\hline
    \multicolumn{5}{c}{MWA-2 robust $+1.0$ tapered}\\\hline
    72--103  & 88 & 12 & 9 &$192 \times 120$ \\
    103--134 & 118 & 6.2 & 9 & $167 \times 105$ \\
    139--170 & 154 & 6.8 & 9 &$181 \times  139$\\
    170--200 & 185 & 7.6 & 9 & $165 \times 130$ \\ 
    200-231 & 216 & 9.2 & 9 & $157 \times 125$ \\\hline
    \multicolumn{5}{c}{MWA-2 robust $0.0$}\\\hline
    139--170 & 154 & 2.6 & 9 &$85 \times  75$\\
    170--200 & 185 & 2.8 & 9 & $70 \times 61$ \\ 
    200-231 & 216 & 2.9 & 9 & $61 \times 53$ \\\hline
    \multicolumn{5}{c}{MWA-2 robust $+1.0$}\\\hline
    170--231$^\text{a}$ & 200 &  2.1 & - & $91 \times 79$ \\\hline
    \end{tabular}\\
    {\footnotesize \textit{Notes.} $^\text{a}$ Only used to create lower right panel in Fig.~\ref{fig:mwa2}.}
\end{table}

The A3391-95 system has been observed with the Murchison Widefield Array Phase 2 \citep[hereafter MWA-2;][]{tgb+13,wtt+18} in its `extended' configuration. The observations cover five frequencies: 88, 118, 154, 185, and 216 MHz, each with 30~MHz bandwidth.  Full processing details for this form of MWA-2 data are described by \citet{Duchesne2020}, but we briefly describe this process here. Observations for the MWA-2 are taken in two-minute snapshot observing mode, with each snapshot calibrated and imaged independently; imaged snapshots are linearly stacked to create deep mosaic images. Calibration is performed using a sky model generated from the GLEAM extra-galactic catalogue \citep{2017MNRAS.464.1146H} using the full-Jones \texttt{Mitchcal} algorithm as described by \citet{oth+16} and full-embedded element primary beam model \citep{scs+17}. For this work, imaging is performed with `Briggs' weighting with robust parameters between $+0.5$ and $+1.0$ with additional Gaussian tapering applied to the 154-, 185-, and 216-MHz datasets to ensure a consistent PSF and $u$--$v$ sampling across the full MWA band. We use the multi-scale CLEAN algorithm within the widefield imager \texttt{WSClean} \citep[][]{wsclean1,wsclean2} for deconvolution. We make a wideband ($\Delta\nu = 60$~MHz) image centred on 200~MHz by stacking the 185- and 216-MHz mosaics made with robust $0.0$ weighting with no tapering as a higher-resolution reference image but note that at this weighting a significant fraction of flux is lost for extended sources. A separate CLEAN component model and residual mosaics are also created and are used for flux-density measurements, with a factor of $\sim0.6$ (dependent on the source size) applied to integrated residual flux density to account for a difference in CLEAN and dirty flux. Image details are presented in Table~\ref{tab:obs:mwa}. GLEAM and MWA-2 images are shown in Fig.~\ref{fig:mwa2}. \par

\section{Results}
\begin{figure}
\center{\includegraphics[width=\columnwidth]{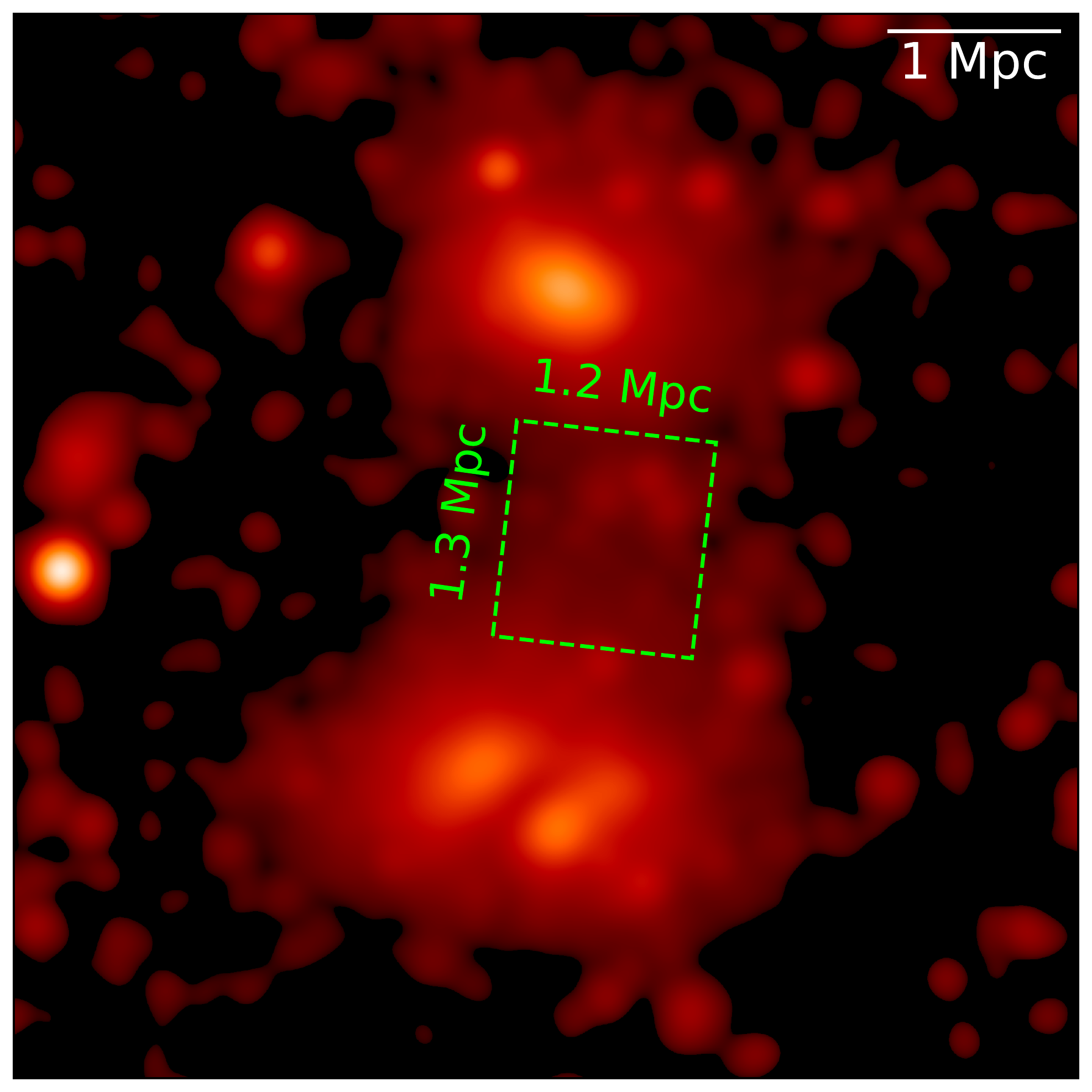}}
\caption{Region of the bridge from the eROSITA X-ray map. The green box denotes the volume we assume for calculating the limits on the emissivity in the bridge region; see Sec. 4.1.}
\label{fig:region}
\end{figure}

\begin{figure*}
\centering
\includegraphics[width=\textwidth]{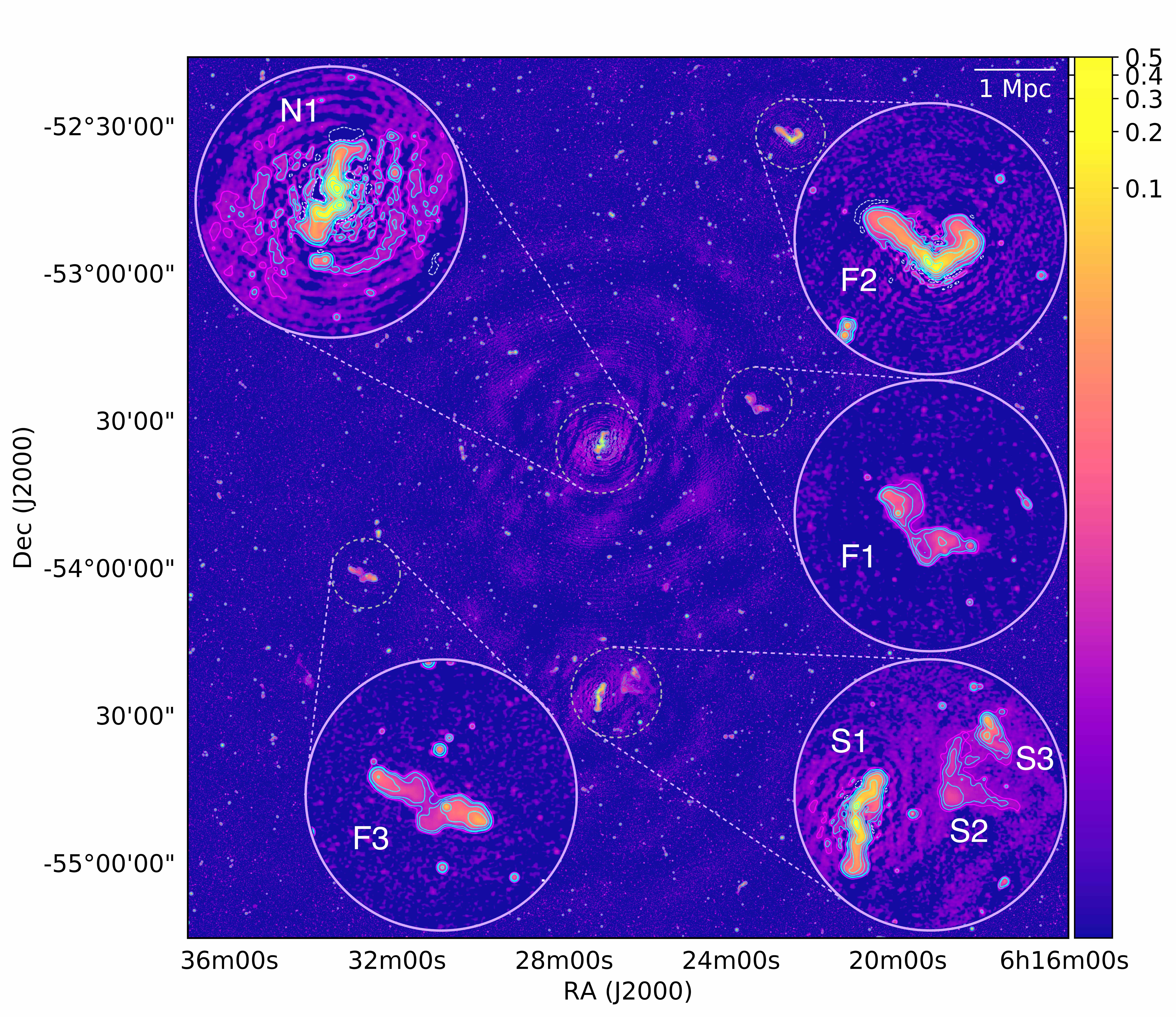}
\caption{Our ASKAP mosaic image of SB8275 covering the Abell 3391-95 system and surrounding sources after direction-dependent calibration. Magenta contours are at $4\sigma$ and cyan contours are $6\sigma \times l$ where $l$ is [1, 2, 8, 32, 128, 512, 1024, 2048] and $\sigma = 36\,\mu$Jy beam$^{-1}$. White dashed contours are $-4\sigma$. All sources highlighted and magnified in this image fall within the mean redshift of the cluster system except for F3  which is at $z=0.193$. Details on these sources are summarised in Tables 2 and 3.}
\label{fig:ASKAPradiomap}
\end{figure*}

\begin{figure*}
    \centering
    \includegraphics[width=1\linewidth]{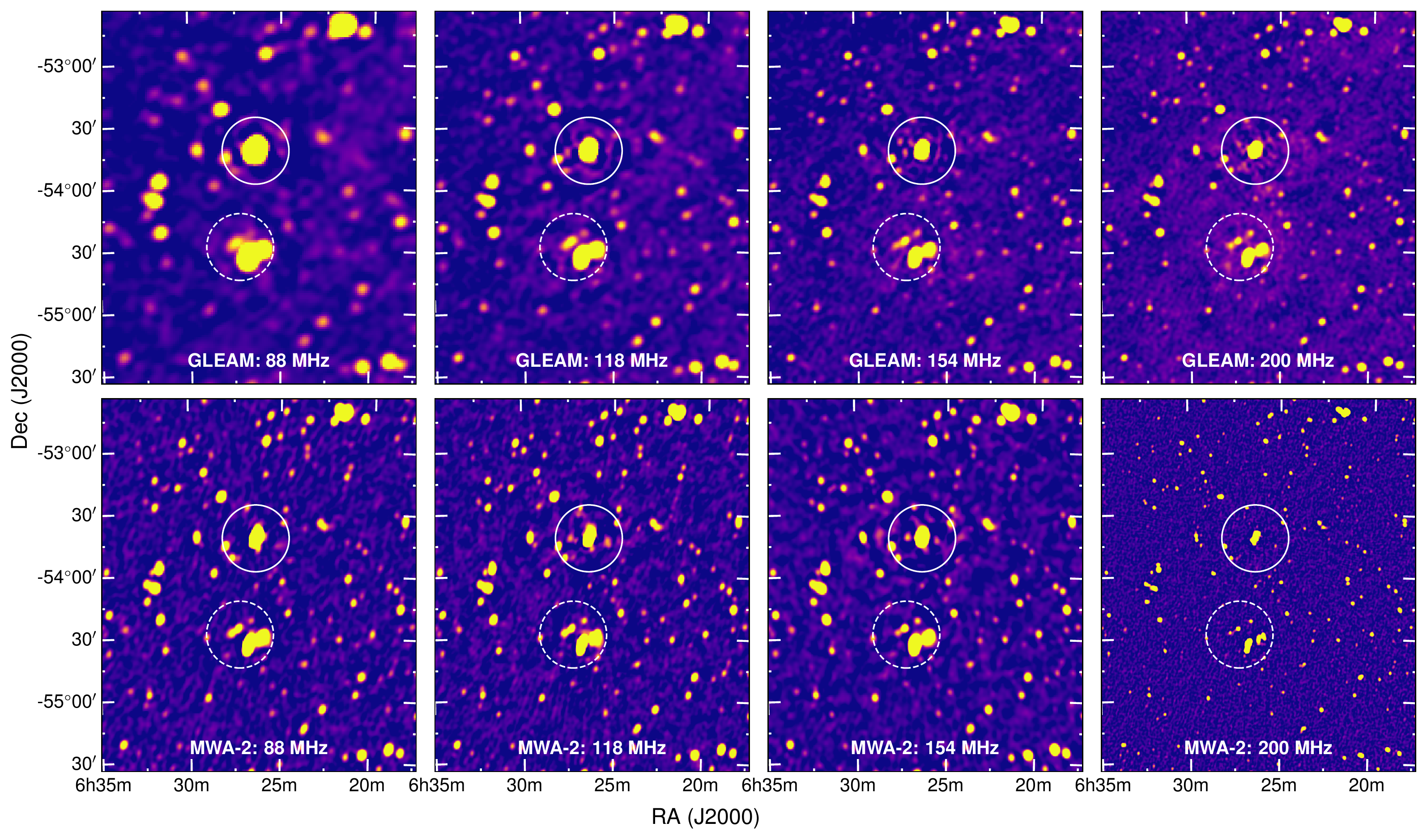}
    \caption{\label{fig:mwa2} GLEAM (top row) and MWA-2 (bottom row) images of the A3391-95 system. The  solid white circle indicates the location of A3391 and the  dashed white circle indicates A3395. Both circles have a 1~Mpc radius at the mean cluster redshift. We note that the colour scale on each map is linear, ranging from $-1\sigma_\text{rms}$ to $20\sigma_\text{rms}$.}
\end{figure*}

\label{sec:results}

In the ASKAP image (Fig.~\ref{fig:ASKAPradiomap}), a radio bridge is not detected, although the short spacings allow us to retain sensitivity up to scales of $\sim$1 degree, significantly larger than the X-ray bridge. In our image after direction-dependent calibration, artefacts on the largest scales as well as negative bowl artefacts are reduced. The noise in the image ranges from $\sim20~\mu$Jy to $\sim50~\mu$Jy per beam where the strongest artifacts emanate from the brightest cluster galaxies at the centre of Abell 3391 and Abell 3395. There does not appear to be any radio emission coincident with the bridge detected by eROSITA. 

No radio bridge is seen in the MWA-2 or GLEAM data (Fig. \ref{fig:mwa2}). We note that the GLEAM data may show faint large-scale emission around bright, complex sources, but this is due to residual un-deconvolved
sidelobes of the PSF characteristic of MWA data. As with the ASKAP data, the bright, complex radio galaxies in each cluster limit the sensitivity that can be achieved in this region. We note that MWA-2 and GLEAM data, despite the expected steep spectra of the bridge synchrotron emission, may not achieve better surface brightness sensitivity than ASKAP \citep[][though we note that for their analysis, MWA-2 are imaged at robust $0.0$, reducing the surface brightness sensitivity for larger scales compared to robust $+1.0$ used here]{Hodgson2020}.

 At the same time, the merging cluster system contains several nearby radio galaxies with extended features that may be interacting with the intracluster environments.  In the following sections we report on the upper limit of the non-detection of the radio bridge and report flux densities for radio galaxies within the cluster field.

\subsection{Upper limit on the radio bridge}

 There seems to be a consensus that the following physical conditions need to be met in order for a radio-synchrotron emission bridge to exist: (1) seed electrons (either from earlier structure formation shocks or from AGN), (2) a magnetic field, (3) an ongoing acceleration mechanism as described in Sect. 1, and finally (4) sufficient time for the acceleration to have produced a sufficient number of electrons of the required Lorentz factors.

In order to compute the flux values in the intercluster region, we placed a square box of 1120 arcsec per side on the bridge region and then compared it to 36 other such boxes outside of the bridge region.
We then blanked areas with flux densities above 0.4 mJy and calculated the mean and rms in the boxes.  Subsequently, we subtracted the mean of the 36 boxes outside of the bridge from the mean of the box on the bridge, and calculated the scatter in the means between the boxes (an appropriate measure of the uncertainty of the flux in each box). The mean (residual) flux in the bridge box is then 39 mJy, and the scatter between boxes is 25 mJy, meaning a non-detection.
The 25 mJy scatter between boxes is approximately 7.5 times larger than the fluctuations on a smaller scale caused by large-scale ripples in the image.

The resulting upper limit on the emissivity is given by $\langle J \rangle_{1\,{\rm GHz}} =4\pi d_L^2 S/V$, where $V$ is the volume of the emitting region and $d_L$ the luminosity distance, which is $d_L\sim 237$ Mpc. For $S$ we take the $3\sigma$, that is, 75 mJy. The boundary of the bridge is difficult to define. If we assume that the bridge roughly follows the emission detected by eROSITA, the radius of the bridge is about 600 kpc or $10'$, and its length is about 1.3 Mpc or $21'$ (see Fig.~\ref{fig:region}). If we assume a cylindrical volume of dimensions 1.3 Mpc $\times \pi\, (0.6$ Mpc)$^2$, we obtain

\begin{eqnarray}
\langle J \rangle_{1\,{\rm GHz}} &<& \frac{ 7.5\times 10^{-28} {\rm W}\,{\rm Hz}^{-1}{\rm m}^{-2}\times 4\pi\times 5.4\times 10^{49}{\rm m}^{2}}{4.3\times 10^{67} {\rm m}^{-3}} \nonumber \\
&<& 1.2 \times 10^{-44} {\rm W}\, {\rm Hz}^{-1} {\rm m}^{-3} .
\end{eqnarray}

If we take $\alpha\sim -1.3$ as in \cite{2019Sci...364..981G}, one would expect an emissivity at 140 MHz of $\langle J \rangle_{140\,\text{MHz}}=1.5\times10^{-43}$~W~Hz$^{-1}$m$^{-3}$, which is a factor 1.7 higher than the emissivity derived for the A399-A401 system. 

The reasons for a non-detection of synchrotron emission from the bridge region at the level of A399-401, for example, may lie in the absence of any one of the four points mentioned at the beginning of this section. If we assume that the relation between thermal gas density and magnetic field strength found in galaxy clusters is also valid in the intercluster region, we can estimate the magnetic field strength using the observed scaling $B\propto n_e
^{0.5}$ \citep{2020ApJ...888..101J}. In cluster centres with electron number densities of $n_e\sim 0.1$ cm$^{-3}$, typical field values of 1-10 $\mu$G are observed. Consequently, the thermal gas density in the bridge region of A3395-A3391 is $n_e\sim 1.5 \times 10^{-4}$ cm$^{-3}$ as inferred from SZ-observations \citep{2017PASJ...69...93S}, which would correspond to a magnetic field strength of $B\sim 0.03-0.3\,\mu$G. We note that a similar value for the electron density, $n_e= 1.08 \times 10^{-4}$ cm$^{-3}$ , was inferred from X-ray observations with Chandra and XMM-Newton by \cite{2018ApJ...858...44A}.
In comparison, the electron density inferred for the intercluster filament in the A399-A401 system is $n_e\sim 2.5-3.3 \times 10^{-4}$ cm$^{-3}$, which is very close to the estimate for A3395-A3391 \citep{2017A&A...606A...1A}. Also, $B=0.1-0.2\, \mu$G was found by \cite{2019Sci...364..981G} in their simulations of A399-401 and $B=0.5\, \mu$G was found by \cite{2020PhRvL.124e1101B} including dynamo amplification. 

From the limit on the surface brightness of $I_0\sim 0.3\,\mu$Jy arcsec$^{-2}$, one can derive an upper limit on the equipartition magnetic field $B_{\rm eq}=(24\pi u_{\rm min}/7)^{1/2}$, where \citep{2004IJMPD..13.1549G}
\begin{eqnarray}
u_{\rm min} & \sim & 2.7\times10^{-11}(1+k)^{4/7}(\nu/\mathrm{GHz})^{-4\alpha/7} \nonumber\\
 & &(I_0/\mathrm{(mJy/arcsec}^2))^{4/7} (d/\mathrm{kpc})^{-4/7}\mathrm{erg\, cm}^{-3} .
\end{eqnarray}
For $\alpha=-1.3$, a proton-to-electron energy ratio of $k=1$, and a source depth of $d_s=800$ kpc, we get $u_{\rm min}\sim 8.1\times 10^{-14} \mathrm{erg\, cm}^{-3}$, and consequently  $B_{\rm eq} = 0.9\, \mu$G. If we were to detect the bridge at the level of our upper limit then the equivalent equipartition field would be 0.9 $\mu$G.  This is higher than derived from our density-scaling argument, and so it is possible that the magnetic fields are simply too low to be detected.

However, we note that the electron density for A399-401 is only slightly higher, $n_e=2.5-3 \times 10^{-4}$ cm$^{-3}$, but the observed surface brightness of its bridge requires much higher fields than the scaled values.

\cite{2020PhRvL.124e1101B} suggest that radio bridges originate from second-order Fermi acceleration of electrons interacting with turbulence. 
The complex dynamics of substructures associated with massive filaments connecting cluster pairs can  generate turbulence even before the two clusters 
undergo their central merger.
Specifically, they explored the role of solenoidal turbulence using a mechanism proposed by \cite{2016MNRAS.458.2584B} where particles interact with magnetic field lines diffusing into super-Alfvenic incompressible turbulent flows.

Provided that the acceleration mechanism in A3391-A3395 is Fermi-II acceleration, the non-detection would imply that the average acceleration time for electrons is longer than the average cooling time; as shown by \cite{2020PhRvL.124e1101B}, this is independent of the magnetic field in the bridge region.
The implication would be that either the turbulent injection scale is larger in A3391-A3395 than in A399-A401 or the turbulent velocity is smaller. We consider the latter to be the most likely because the turbulent injection length should be similar for systems of similar sizes while the turbulent velocity can vary significantly.
A smaller turbulent velocity in A3391-A3395 could be motivated by the smaller mass of the system and by the fact that these two systems are separated by a greater distance (in terms of virial radius of the clusters) than that separating A399-A401, possibly meaning that the merger is in a less advanced stage.
Also, the A3391-A3395 clusters are less massive than either A1758 or A399-A401.
Furthermore, as also briefly mentioned in paper 1, we cannot know the true three-dimensional distance between the systems. It is possible that A3391 and 3395 are much further apart than what is implied by their projection, as is indeed suggested by the X-ray analysis in Reiprich et al. (2020). This would likely result in lower estimates of electron densities, and might also explain the absence of giant radio halos in the constituent clusters, as the occurrence of giant radio halos depends strongly on mass \citep{2010A&A...517A..10C}.
Given their similarity to A399-A401 in other aspects (both premerging, similar gas densities in the bridge region), the A3391-A3395 system is a promising target for deeper radio observations.  

For a more detailed study of the implications of the non-detection, dedicated MHD simulations are needed, such as those recently presented in Locatelli et al. (2020). These latter infer upper limits in LOFAR observations of two cluster pairs, RXCJ1659.7+3236-RXCJ1702.7+3403 and RXCJ1155.3+2324-RXCJ1156.9+2415, yielding upper limits on the magnetic field in the intercluster region of $B<0.2\, \mu$G. Using cosmological simulations, these latter authors infer primordial magnetic fields of less than 10 nG.

\subsection{Other notable radio features}

In addition to the absence of diffuse emission in the intercluster bridges, the environment of this unusual cluster-pair field shows a number of very interesting sources that are worth cataloguing given the potential impact of the cluster system. By no means is this a complete sample, as it was compiled by identifying extended objects by eye. The field contains a number of candidates for diffuse radio emission that will need to be followed up with deeper X-ray and radio observations. In addition, the field contains a number of Giant Radio Galaxies (GRGs) of which all but one lie in the background of this cluster pair. The exception is the FR\,I radio galaxy J0621--5647 in the far south, at the same $z$ as the
cluster pair. Polarisation measurements of these GRGs could be interesting for rotation-measure studies of the magnetic fields in the double cluster region \citep{2013MNRAS.432..243P,2020A&A...638A..48S}. Finally, there are some other notable radio galaxies that show signs of interaction with the ICM of A3391-A3395, such as the wide-angle tail source F2 as well as the radio galaxy F1. The latter shows lobes that are transversally advected, which may be caused by the galaxy falling into the cluster (see Fig.~\ref{fig:overlay}).\\

Table~\ref{tab:flux} lists the measurements of flux density, $S$, and spectral index, $\alpha$,  for radio sources in the A3391-A3395 cluster field. Flux density was measured within a region marking the 3$\sigma$ contour line using a local rms value for $\sigma$. To measure the spectral index of specific sources or source regions (except for S3), we convolved the ASKAP-DD image to the same resolution as our tapered MWA-2 observations. For S3, to ensure S2 and S3 were resolved enough for flux-density measurements, we use the native resolution ASKAP-DD image and the robust 0.0 MWA-2 images. We do not measure S2 directly with the MWA-2 data as the diffuse emission becomes resolved out in the robust 0.0 MWA-2 images. We used \href{https://gist.github.com/Sunmish/198ef88e1815d9ba66c0f3ef3b18f74c}{fluxtools.py} to measure the flux density and error on the flux density when considering the rms. The uncertainties are calculated taking into account the image noise and flux-scale errors as follows,

\begin{equation}
 \Delta S = \sqrt{\Delta S^2_\text{image} + \Delta S^2_\text{fluxscale} }= \sqrt{N\times \sigma^2 + (f \times S)^2},
\end{equation}
where $N$ is the number of independent beams covering the source, $f=10\%$ is the flux-scale error for the EMU observation, and $f=9\%$ for the MWA-2 observations. The spectral index is estimated by fitting the flux densities of the sources to a power law as a function of the observed frequency.\\

\begin{table*}
\caption{Properties of radio galaxies in Fig.~\ref{fig:ASKAPradiomap} and their respective host galaxies.  Columns are (1) a shortened name where `EMU ES' stands for Evolutionary Map or the Universe Early Science Source, (2) the source label according
to Figs.~\ref{fig:overlay} and \ref{fig:ASKAPradiomap}, (3) the name of the optical host galaxy, (4) its
spectroscopic redshift, (5) its largest angular size (LAS) measured
in a straight line between opposite ends of the detectable radio emission,
(6) its largest projected linear size (LLS) assuming the cosmological parameters
listed at the end of Sect. 1.}
\begin{tabular}{ccccccccc}
\hline
 EMU ES & Our label & Host name & Redshift & \multicolumn{2}{c}{Extent of radio lobes} \\
 name & (in Fig.~\ref{fig:ASKAPradiomap})&  & $z$ & LAS & LLS  & Comments \\
 & & & & [arcmin] & [kpc] \\
\hline
J0626--5341 & N1 & ESO 161-IG 007 NED02 & 0.0551 &  5.35 & 346 & FR\,I in A3391 \\
J0626--5433 & S1 & WISEA J062649.57--543234.4 & 0.0520 &  6.07 & 372 & PKS\,0625--545, in A3395 \\
J0626--5432 & S2 & -- ? & -- &  5.52 & -- & relic/remnant in A3395 \\
 J0625--5427 & S3 & WISEA J062557.04--542750.4 ? & 0.0603 &  3.25 & 229 & remnant/relic in A3395 \\
J0622--5334 & F1 & WISEA J062255.56--533434.5 & 0.0567 &  5.71 & 379 & asym. FR\,II?, PMN\,0622--5334 \\
J0621--5241 & F2 & 2MASX J06214330--5241333 & 0.0511 &  6.21 & 374 &  WAT, PKS\,0620--52 \\
J0632--5404 & F3 & WISEA J063201.16--540457.4 & 0.193 & 5.80 & 1117 & FR\,II plume \\
\hline 
\end{tabular}
\label{tab:RGs_EMU}
\end{table*}

\begin{table*}
\centering
\caption{\label{tab:flux} Integrated flux density measurements (within $3\sigma$ contours where $\sigma$ is the local rms) of sources labelled in Fig. \ref{fig:ASKAPradiomap}.}
\begin{tabular}{c | c c | c c c c c c }\hline
Source & ASKAP-DD & rms ($\sigma$) & MWA-2 & MWA-2 & MWA-2 & MWA-2 & MWA-2 & ASKAP-DD$^\text{a}$  \\\hline
      & 1013 MHz & local & 88 MHz & 118 MHz & 154 MHz & 185 MHz & 216 MHz  & 1013 MHz  \\
      & [mJy] & [$\mu$Jy/beam] & [Jy] & [Jy] & [Jy] & [Jy] & [Jy]  & [Jy]  \\ \hline     
N1 & $8799 \pm 880$ & 50 & $89.6 \pm 9.7$ & $62.1 \pm 6.2$ & $48.6 \pm 4.3$ & $41.1 \pm 3.7$ & $35.3 \pm 3.2$ & $9.06 \pm 0.91$  \\
S1 & $4314 \pm 431$ & 50 & $33.2 \pm 4.0$ & $24.5 \pm 2.7$ & $19.8 \pm 1.9$ & $17.2 \pm 1.6$ & $15.2 \pm 1.6$ & $4.41 \pm 0.44$  \\
S2 & $72.1 \pm 7.2$ & 48 & - & - & $1.38\pm0.29$ $^{c}$ & $1.17\pm0.23$ $^{c}$ & $0.91\pm0.20$ $^{c}$ & -  \\
S3 & $226 \pm 22.6$ & 48 & - & - & $1.26\pm0.12$ & $0.99\pm0.10$ & $0.83\pm0.10$ & -  \\
S2$+$S3$^\text{b}$ & $298 \pm 29.8$ & 48 & $5.92 \pm 0.87$ & $3.81 \pm 0.48$ & $2.64 \pm 0.26$ & $2.15 \pm 0.21$ & $1.73 \pm 0.17$ & $0.392 \pm 0.039$  \\
F1 & $175 \pm 17.5$ & 30 & $0.786 \pm 0.104$ & $0.532 \pm 0.063$ & $0.565 \pm 0.057$ & $0.428 \pm 0.043$ & - & $0.167 \pm 0.017$  \\
F2 & $4883 \pm 488$ & 40 & $25.5 \pm 3.2$ & $18.8 \pm 2.1$ & $15.5 \pm 1.5$ & $13.8 \pm 1.3$ & $11.6 \pm 1.1$ & $4.90 \pm 0.49$  \\
F3 & $556 \pm 55.9$ & 26 & $3.14 \pm 0.36$ & $2.28 \pm 0.24$ & $1.89 \pm 0.17$ & $1.58 \pm 0.14$ & $1.39 \pm 0.13$ & $0.586 \pm 0.059$  \\\hline
\end{tabular}\\
{\footnotesize \textit{Notes.} $^\text{a}$ Convolved to a resolution of $150$~arcsec $\times$ $131$~arcsec to match the tapered MWA-2 images, which results in an additional contribution from nearby but faint point sources. $^\text{b}$ S2 and S3 are convolved together in the tapered and 88--118 MHz MWA-2 images. $^{c}$ Residual flux density after subtracting S3 from S2$+$S3.}
\end{table*}

\begin{figure*}
\centering
  \includegraphics[width=8cm]{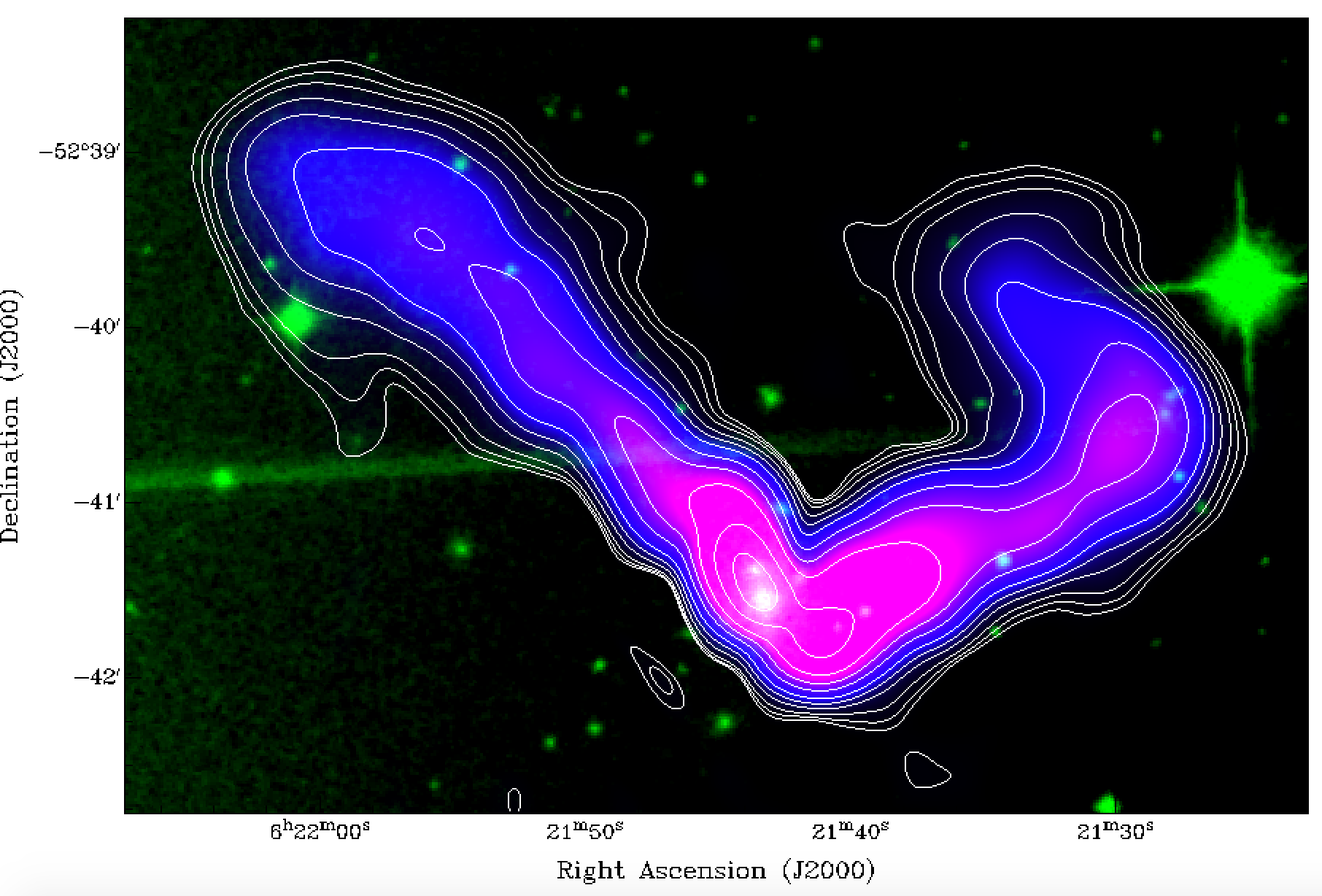}
  \includegraphics[width=8cm]{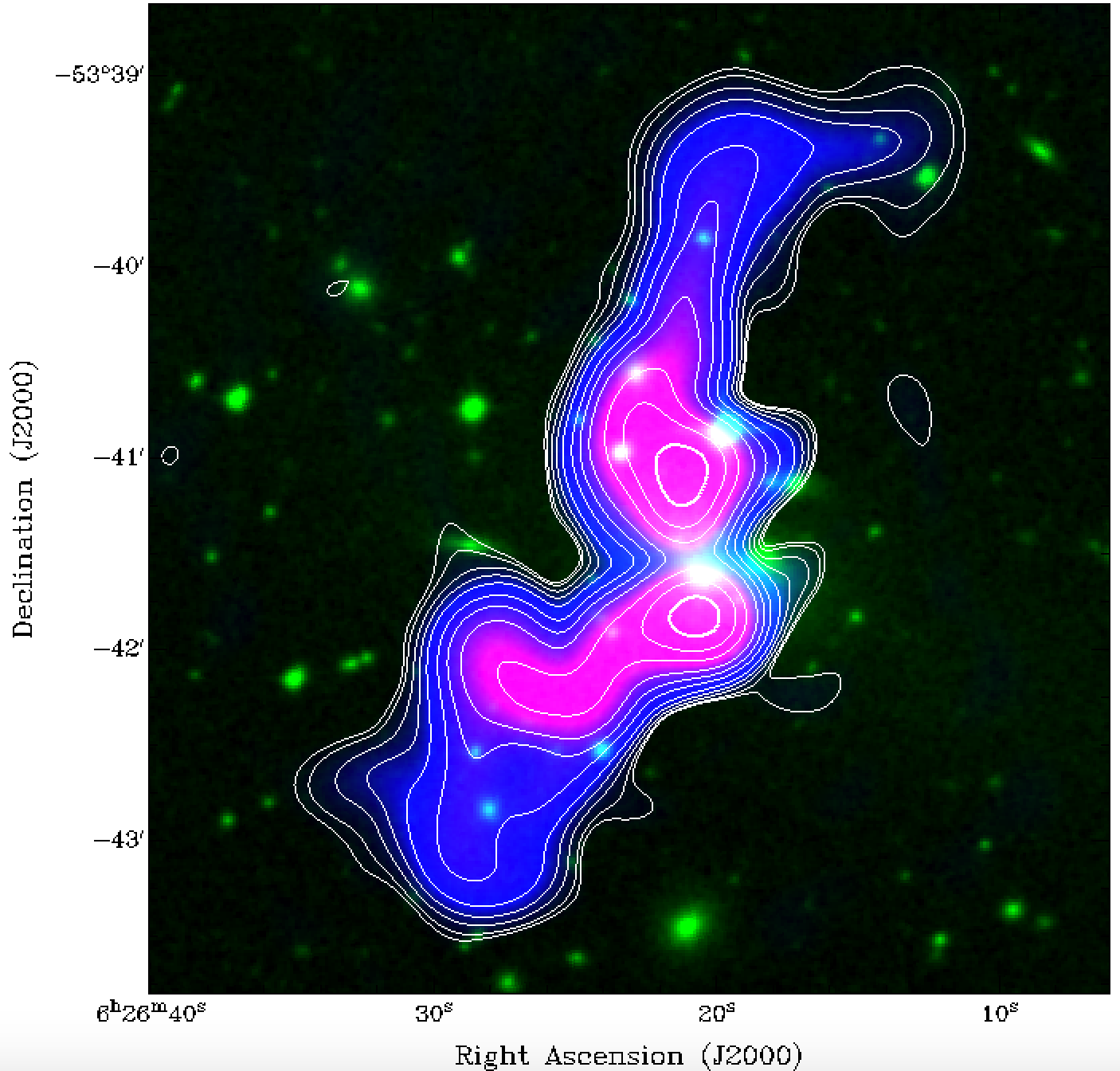}
  \includegraphics[width=8cm]{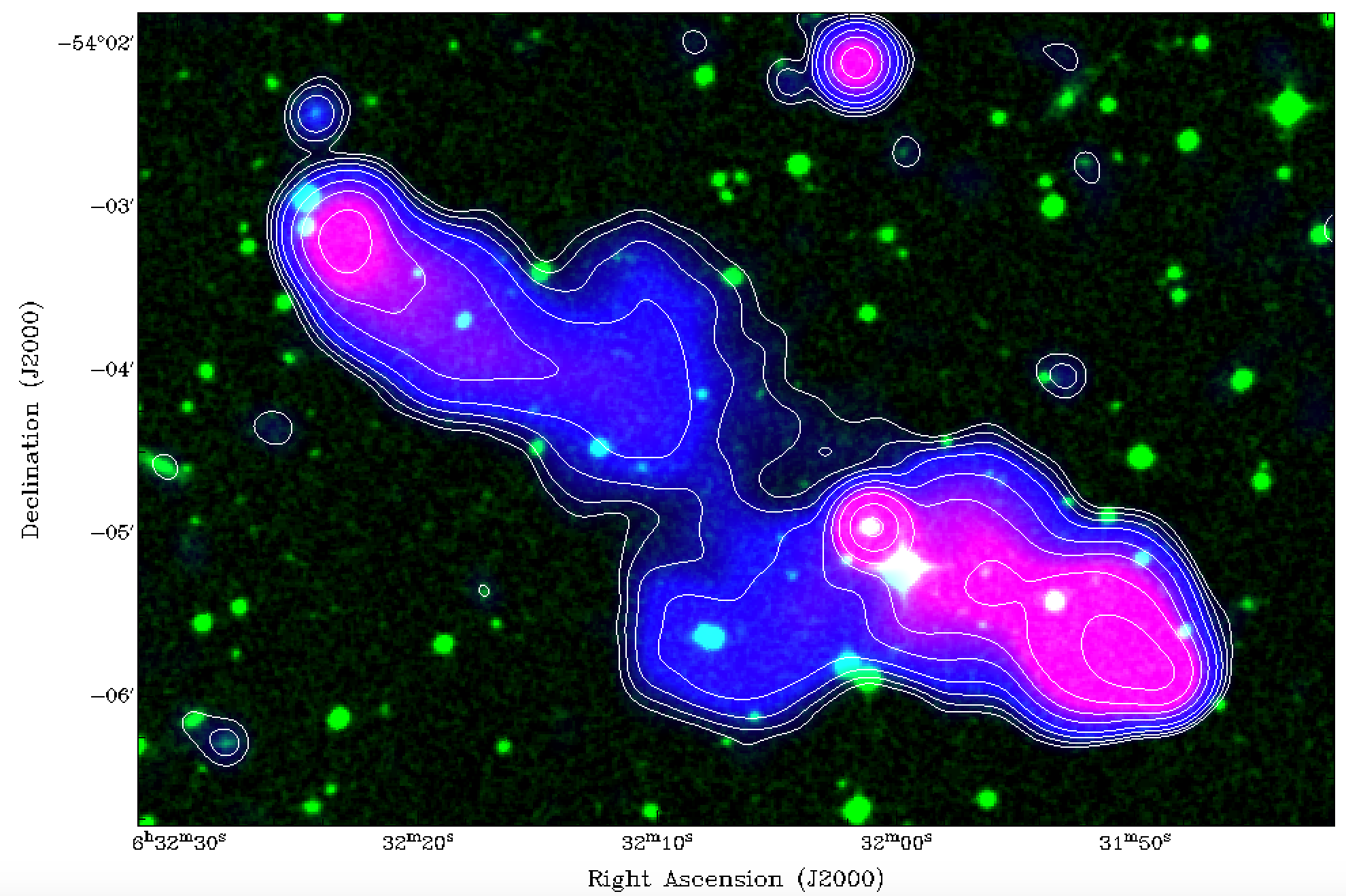}
  \includegraphics[width=8cm]{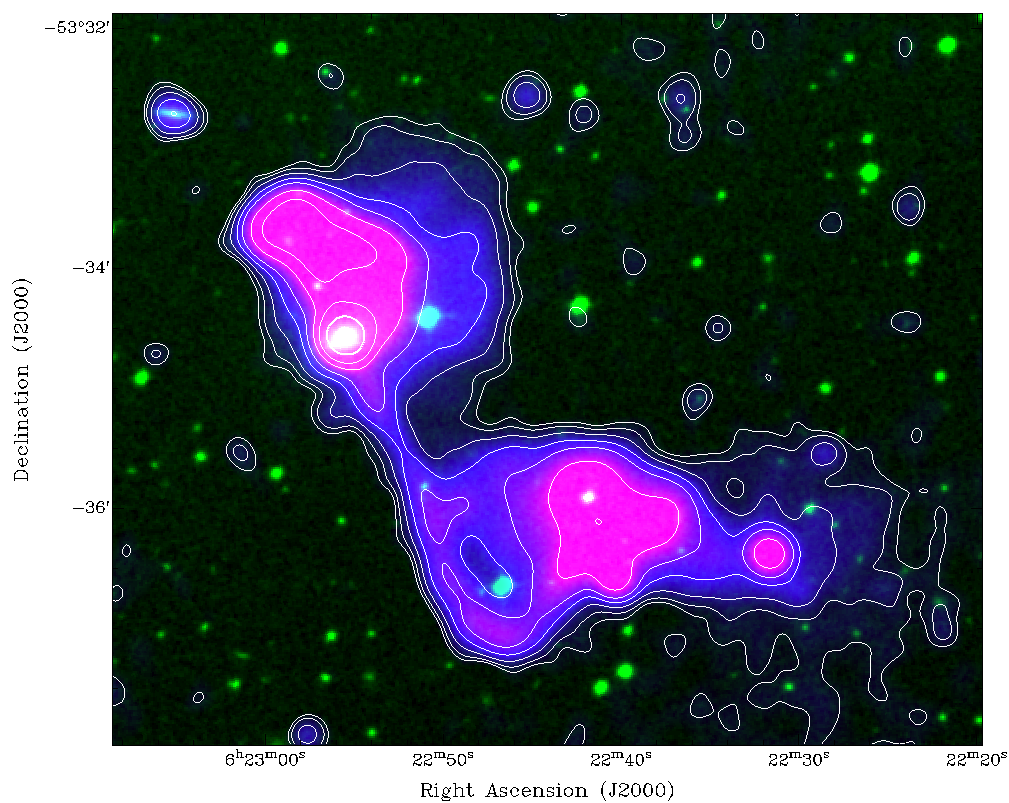}
  \includegraphics[width=8cm]{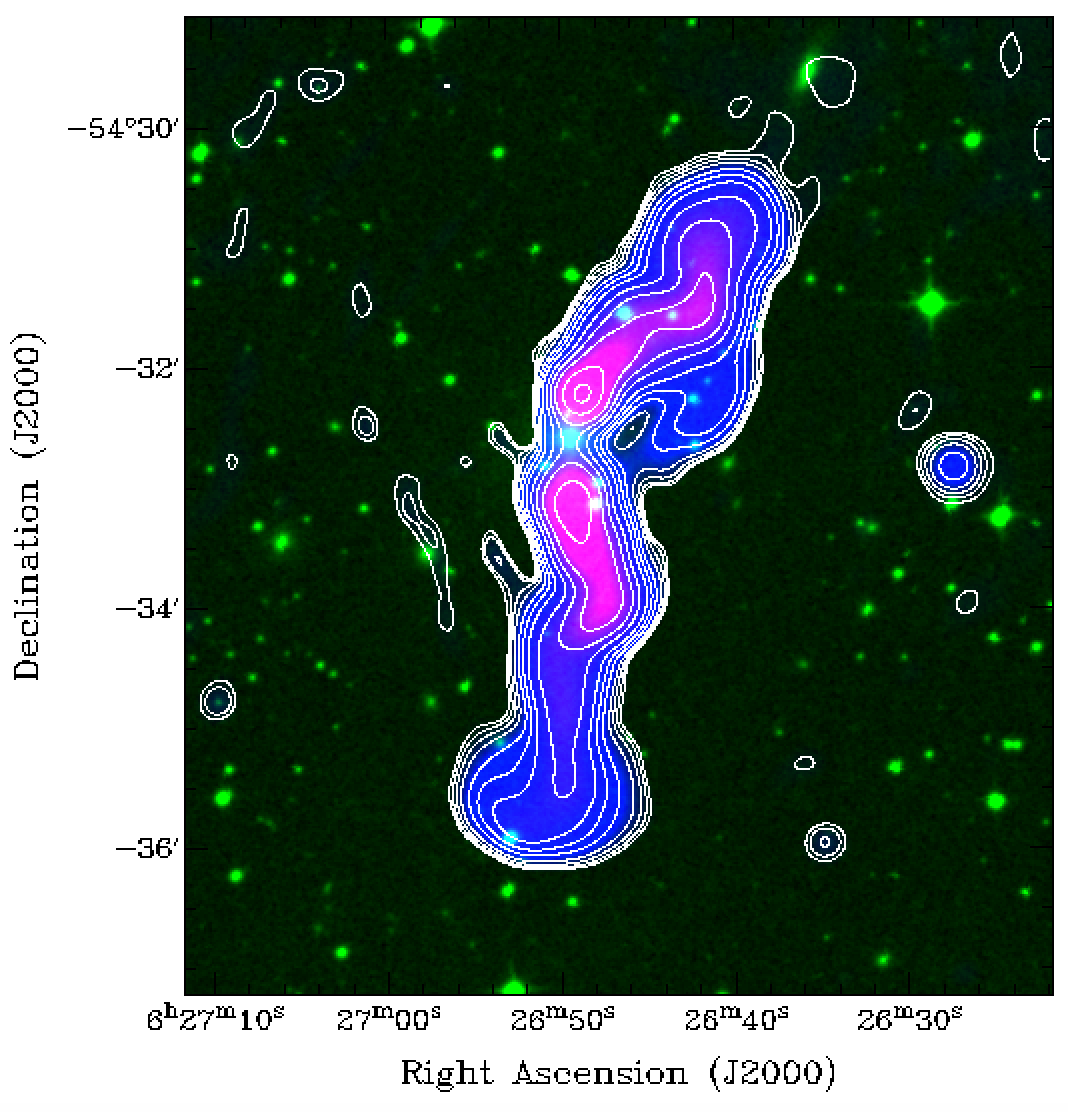}
  \includegraphics[width=8cm]{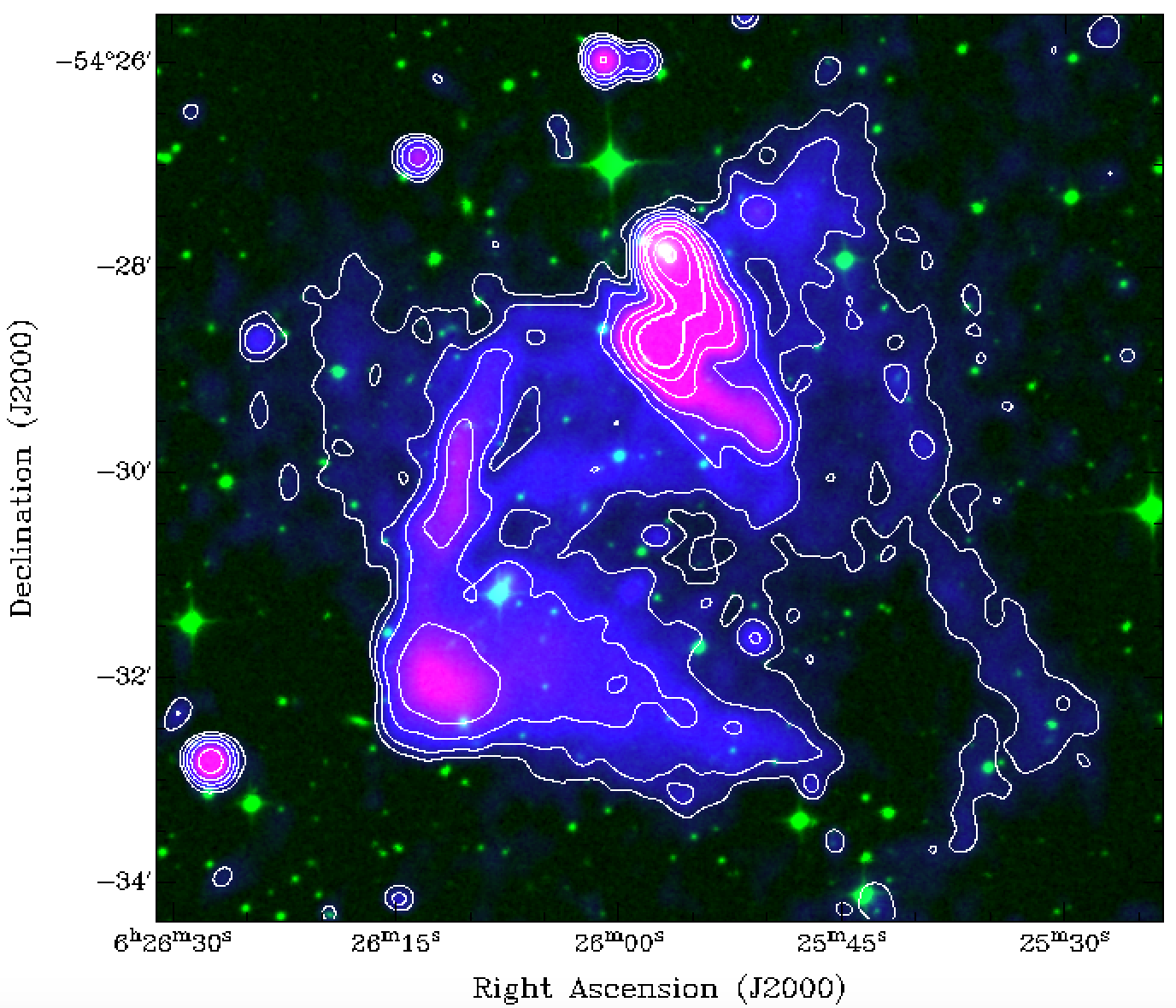}
\caption{ASKAP 1 GHz radio continuum images of the radio galaxies shown in Fig. 1 overlaid on DSS2 $R$-band images. Top left: F2, Top right: N1, Middle left: F3, Middle right: F1, Bottom left: S1, Bottom right: S2 and S3. The radio contour levels starting at 3$\sigma$ are typically 0.012, 0.025, 0.05, 0.1, 0.25, 0.5, 1, 2.5, 5, 10, 25, 50, 100, 250 and 500 mJy/beam, except for F1-F3 where a 0.006 mJy/beam contour is added, and N1 where the 0.012 mJy/beam contour is omitted.}
\label{fig:mainsource}
\end{figure*}

\subsubsection{Source N1 (EMU ES J0626--5341)}
The bright radio galaxy   EMU ES J0626--5341 (EMU ES: Evolutionary Map of the Universe Early Science Source) is located in the centre of the Abell~3391 cluster and extends over 4.5 arcmin (see Fig.~\ref{fig:mainsource}). It is also located at the centre of the X-ray emission (see Fig.~\ref{fig:xrayover}). Its host is the elliptical galaxy 2MASX J06262045--5341358 ($PA$ = 54\degr), which is  the eastern component of ESO 161-G008, with a redshift of $z$ = 0.055. We estimate a linear projected radio size of 290 kpc. The radio lobes show numerous twists possibly due to density variations in the ICM. 

\subsubsection{Source S1 (EMU ES J0626--5432)}
\begin{figure*}
\includegraphics[width=\textwidth]{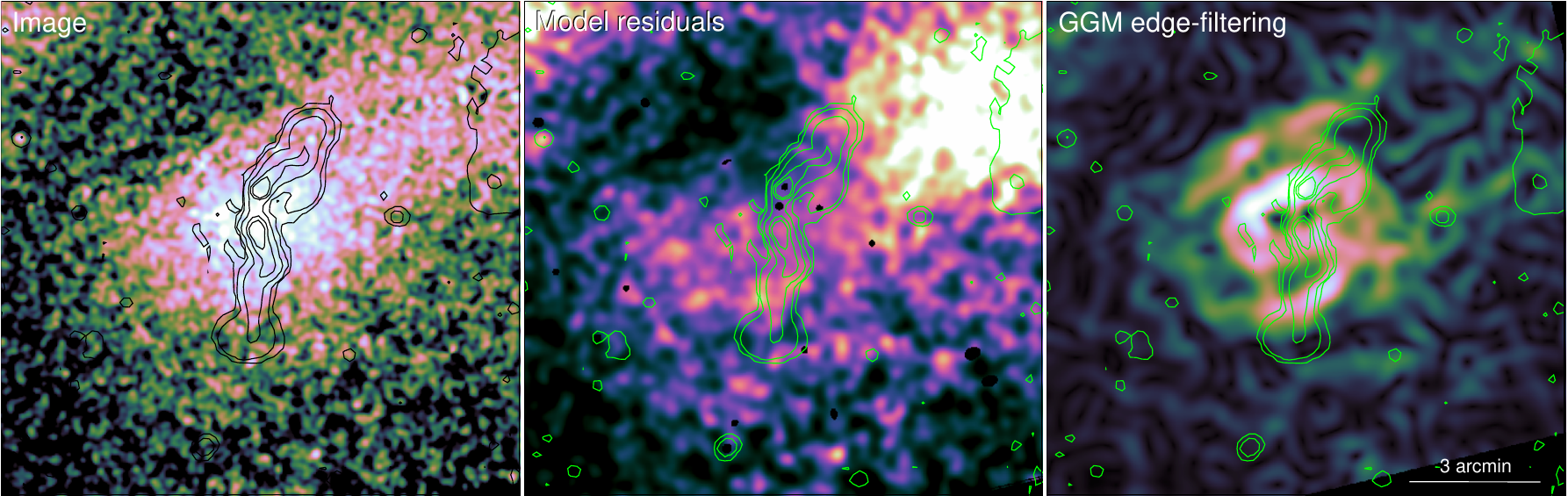}
\caption{Chandra X-ray image of the S1 region and radio contours. Left: Exposure-corrected Chandra image in the 0.5-5 keV band after masking point sources, smoothed by a Gaussian of $\sigma=3.9$~arcsec. The contours show the EMU data at levels of $10^{-4}$, $10^{-3}$, $0.01$, $0.05$ and $0.1$ Jy~beam$^{-1}$. Centre:  Fractional difference between an exposure-corrected image smoothed by a Gaussian with $\sigma=7.9$~arcsec and a smooth model which is its average at each radius (measured from the radio nucleus). Right: Edge-filtered X-ray image showing the gradient magnitude on scales of $16$~arcsec, meaning that components on scales larger than $16$~arcsec have been filtered out.}
\label{fig:chandra_s1}
\end{figure*}

\begin{figure}
\includegraphics[width=1.\columnwidth]{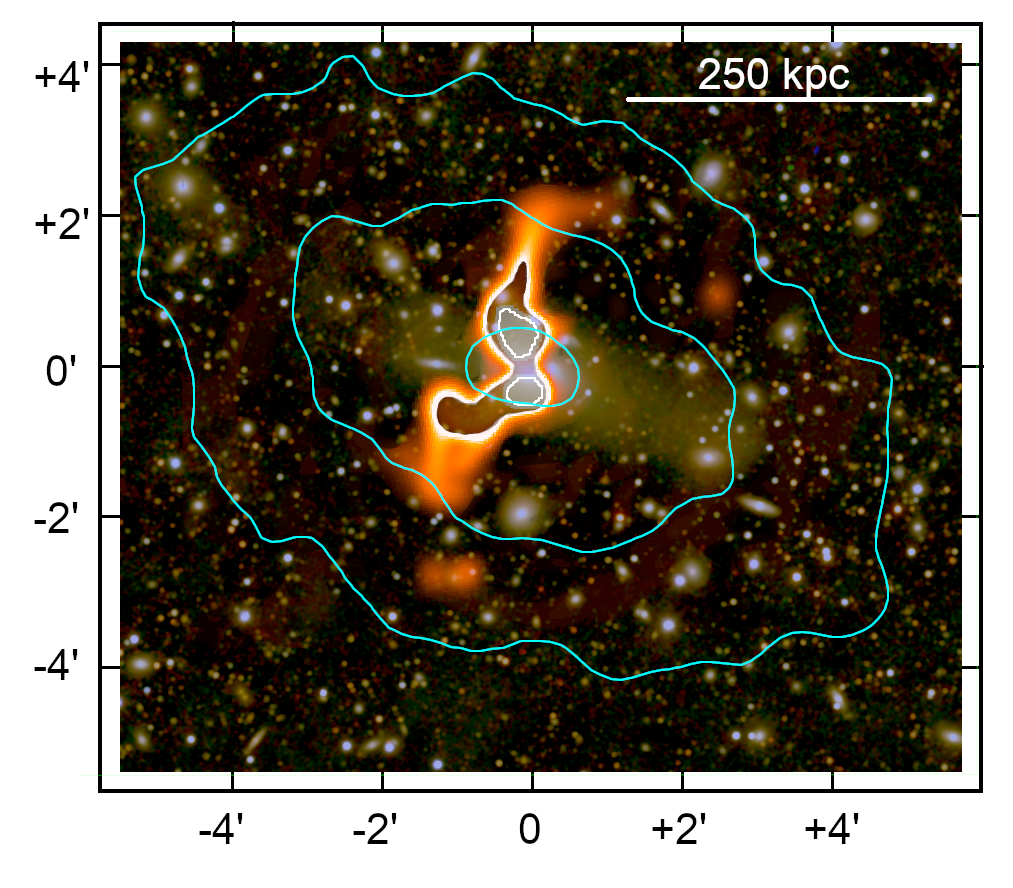}

\includegraphics[width=0.95\columnwidth]{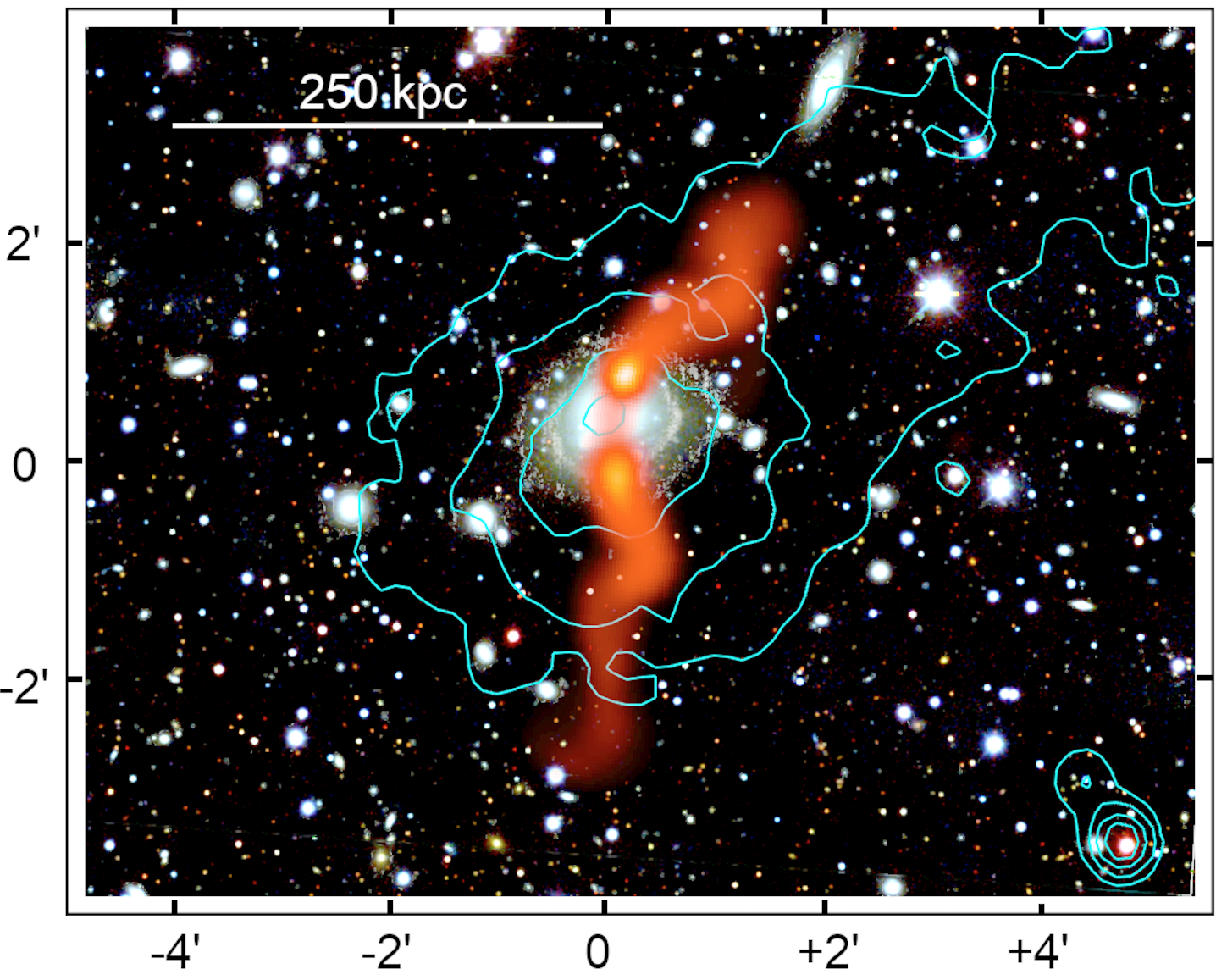}

\includegraphics[width=0.95\columnwidth]{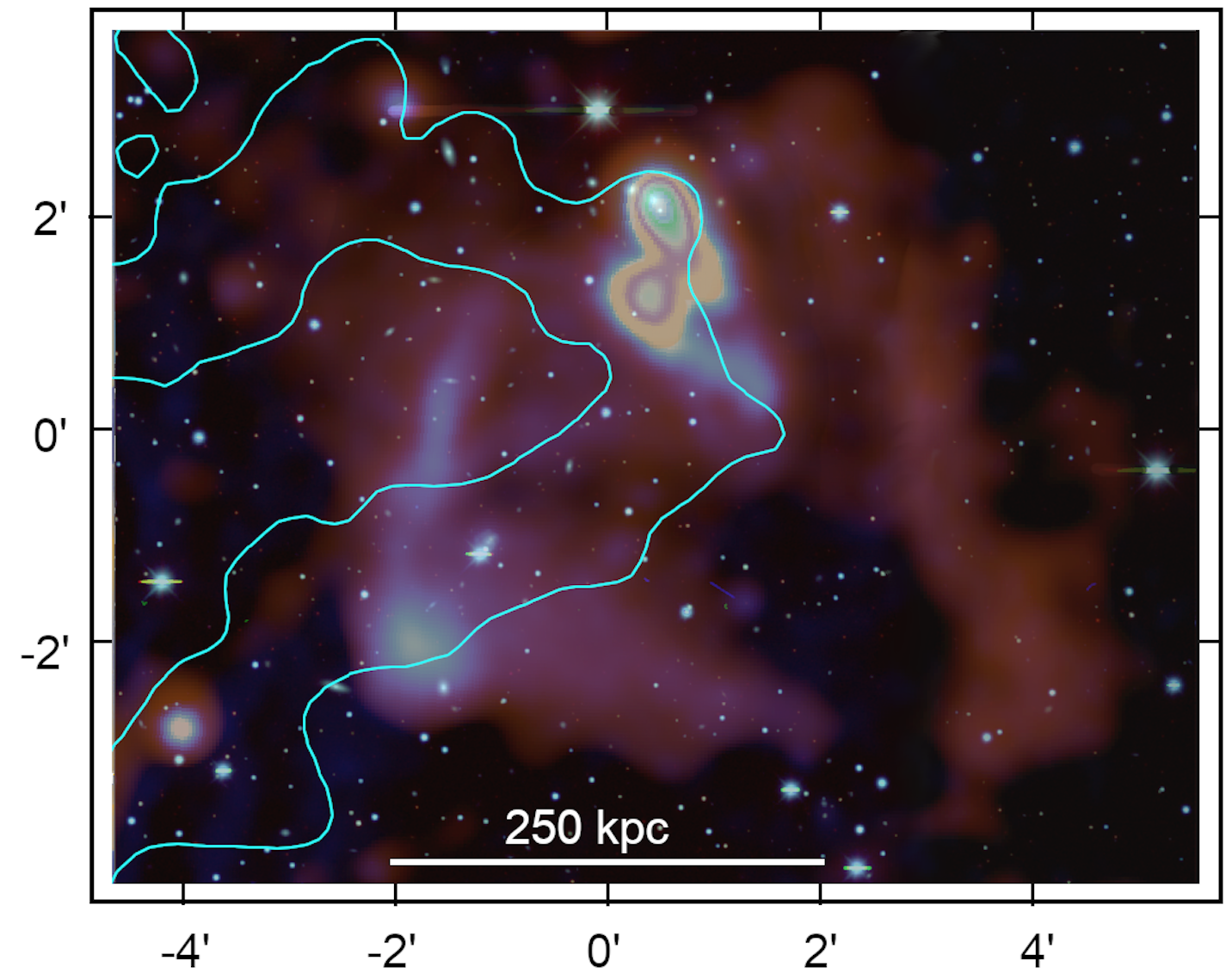}
\caption{Overlays of radio images on optical images from the g, r, and z images from the DECam Legacy Survey. The turquoise contours are from eROSITA. Top:  Multi-wavelength close-up around radio galaxy N1. The radio image from the EMU pilot survey at a resolution of 11" is shown in orange, with the high-brightness areas masked out and replaced with white contours.  Middle: Multi-wavelength close-up around the radio S2/S3 region. The radio emission from the EMU ES observation at a resolution of $11''$ is shown in blue, overlaid in orange with a lower, $25''$ resolution image. The high brightness areas are colour-coded separately. Bottom: Multi-wavelength close-up around radio galaxy S1.  The radio image from the EMU pilot survey at a resolution of $11''$ is shown in orange.}
\label{fig:xrayover}
\end{figure}

The bright radio galaxy EMU ES J0626--5432 (PKS 0625--545) is a Fanaroff-Riley class I (FR I) \citep{1974MNRAS.167P..31F} source located in the centre of the A3395 cluster and extends over 5.5 arcmin (see Fig.~\ref{fig:mainsource}). Its host galaxy is WISEA J062649.57--543234.4 ($z$ = 0.052). Radio fluxes from 145 MHz to 8.5 GHz as well as 148 GHz are listed in NED. The SUMSS 843 MHz flux is catalogued as $\sim$5 Jy and the Australia Telescope Compact Array (ATCA) 1.4 GHz flux as $\sim$3.1 Jy \citep{2011ApJ...743...78L}. To the south, just as it starts to exit the denser X-ray gas (see Fig.~\ref{fig:chandra_s1}), the tip of the radio jet starts to expand spherically and the brightness of the emission drops. This could be the onset of the formation of an X-ray cavity. Comparison with the eROSITA data also shows that the BCG is offset from the centroid of the X-ray emission (see Fig.~\ref{fig:xrayover}). This was studied by \cite{2014ApJ...797...82L, 2020MNRAS.tmp.3079D} for example. In X-ray groups in COSMOS, we showed that BCGs can be offset from the group centre \citep{2019MNRAS.483.3545G}. This has also been studied in HIFLUGCS clusters \citep{2011A&A...526A.105Z}.

The interaction of the radio source with the surrounding ICM can be seen in more detail in a Chandra X-ray image of the region. After reprocessing the Chandra observation (OBSID 4944), images and exposure maps were created, and point sources were masked. Figure~\ref{fig:chandra_s1} shows the smoothed exposure-corrected image, the fractional difference from a model created from the radial average, and an edge-filtered X-ray image made using using the GGM algorithm \citep{2016MNRAS.457...82S, 2016MNRAS.460.1898S}. There is some indication of a depression in the X-ray surface brightness in the region of the round, southern tip, but it is of low significance \citep{2011ApJ...743...78L}. The northern tip appears to end where there is an excess of emission in the smooth model residual map. The edge-filtered map confirms that the two ends of the radio source are located where there are steep gradients in the X-ray surface brightness. There is also a clear connection between the structure within the radio source and the X-ray gradient filtered image.

\subsection{Source S2+S3}

The sources S2 and S3 form a complex conglomerate of radio galaxies and interconnected diffuse emission (see Fig.~\ref{fig:mainsource}). The integrated spectral index between 88 MHz and 1013 MHz is $\sim -1$. Figure~\ref{fig:mainsource} shows an overlay of the S2-S3 complex over the optical DSS2 R-band image.
The L-shaped source S2 is fed by the radio galaxy 2MASX J06261051-5432261 that fans out into two arms towards the north and the west that are roughly perpendicular to one another. The arms extend for about 4 arcmin. There are further faint filaments of emission that stretch out to the source S3. 

From the brightest spot close to the galaxy 2MASX J06255706-5427502, the source S3 extends to the northwest before it turns to form a faint, linear source that extends for about 6 arcmin towards the southwest. This could be a radio relic or phoenix that shows re-acceleration of plasma that may originate from an AGN. Interestingly, the southern part of this linear feature lies in a high-temperature region as indicated by the oxygen-to-softband ratio in the eROSITA data (see Fig. 18 in Reiprich et al. 2020). It is not clear whether this region corresponds to a shock or is merely a region of higher pressure that could boost the synchrotron emission from old radio plasma that may have come from a radio galaxy.

\subsection{Source F1 (EMU ES J0622--5334)}
EMU ES J0622--5334 (PMN\,0622--5334) is a large, rather asymmetric radio galaxy with Fanaroff-Riley class I morphology in Abell S0584 (see Fig.~\ref{fig:mainsource}). Its host galaxy is WISEA J062255.56--533434.5 at $z$ = 0.0567. The radio lobes extend over $\sim$6 arcmin, corresponding to $\sim$400 kpc. The SUMSS 843 MHz flux of PMN\,0622--5334 is catalogued as $75 \pm 6$ mJy, likely not including the low-surface brightness emission. The 4.85 GHz flux is catalogued as $47 \pm 8$ mJy \citep{1996ApJS..103..145W}.

\subsection{Source F2 (EMU ES J0621--5241)}
EMU ES J0621--5241 (PKS\,0620--52) is a large wide-angle tail (WAT) radio galaxy (see Fig.~\ref{fig:mainsource}) which is located far beyond the virial radius of A3391 and could be infalling into the system.
Its host galaxy is 2MASX J06214330--5241333  at $z$ = 0.0511. It is associated with RX J0621.6--5241. The two radio lobes have extents of 4 arcmin (NE; 240 kpc) and 2.5 arcmin (NW; 150 kpc). 

\subsection{Source F3 (EMU ES J0632--5404)}
EMU ES J0632--5404 is a GRG with an FR\,II morphology extending 5.8 arcmin (see Fig.~\ref{fig:mainsource}) along a position angle of $PA \sim 60$\degr. Its host galaxy is the quasi-stellar object (QSO) WISEA J063201.16--540457.4, 6dFGS gJ063201.2--540458, SWIFT J0632.1--5404 for which \cite{1998AJ....115.1253P} measured a redshift of $z=0.193$, while \cite{2005AJ....130..896S} measured $z=0.2036$. Comparing both spectra we believe the former redshift (0.193) to me more reliable, and therefore find an LLS of 1.12 Mpc. \cite{2005AJ....130..896S} measure a 843 MHz flux of 694 mJy.
This source is also GLEAM J063153-540527 with a low-frequency
spectral index of -0.88, and so ASKAP may show more of the extended
emission as the spectrum seems to flatten above 800 MHz.
The 4.85 GHz flux from PMN (source J0631-5405) is $0.155 \pm 0.011$ Jy, resulting in a spectral index of $\alpha^{1013}_{4850}=-0.85$, in agreement with low frequencies.
The host is also listed as blazar BZQ J0632-5404 by \cite{2009A&A...495..691M}.


\subsection{Large and Giant Radio Galaxies}

Based on visual inspection of the whole 30 deg$^2$ field for radio galaxies, we identified close to 200 objects larger than $\sim$1 arcmin. Of these, 27 are GRGs with linear projected sizes greater than 1~Mpc (7 of these are presently still candidate GRGs). Only a single one was already known, namely J0632-5404, published as SGRS J0631--5405 by \cite{2005AJ....130..896S}.
A further 30 radio galaxies not listed in the present paper are larger than 700~kpc (of which 7 are candidate GRGs). 

Giant Radio Galaxies are thought to reside in low-density environments where the jets face little resistance.  However, about 10 percent of GRGs have now been found to reside in cluster environments. According to \cite{2020A&A...642A.153D}, about 820 GRGs larger
than 0.7 Mpc are known to date. However, owing to their large extent and low surface brightness, they are notoriously difficult to detect. Using Data Release 1 of LoTSS, \cite{2020A&A...635A...5D} found 239 GRGs (there, defined as galaxies with LLS $>0.7$ Mpc) in 424 deg$^2$ of sky, albeit in a frequency range of 120--168 MHz. For galaxies with LLS $>0.7$ Mpc, we obtain a sky density of $\approx 1.7$ deg$^{-2}$, while \cite{2020A&A...635A...5D} found 239/424 = 0.55 deg$^{-2}$, three times less than our value.

\begin{figure*}
\center{\includegraphics[width=\textwidth]{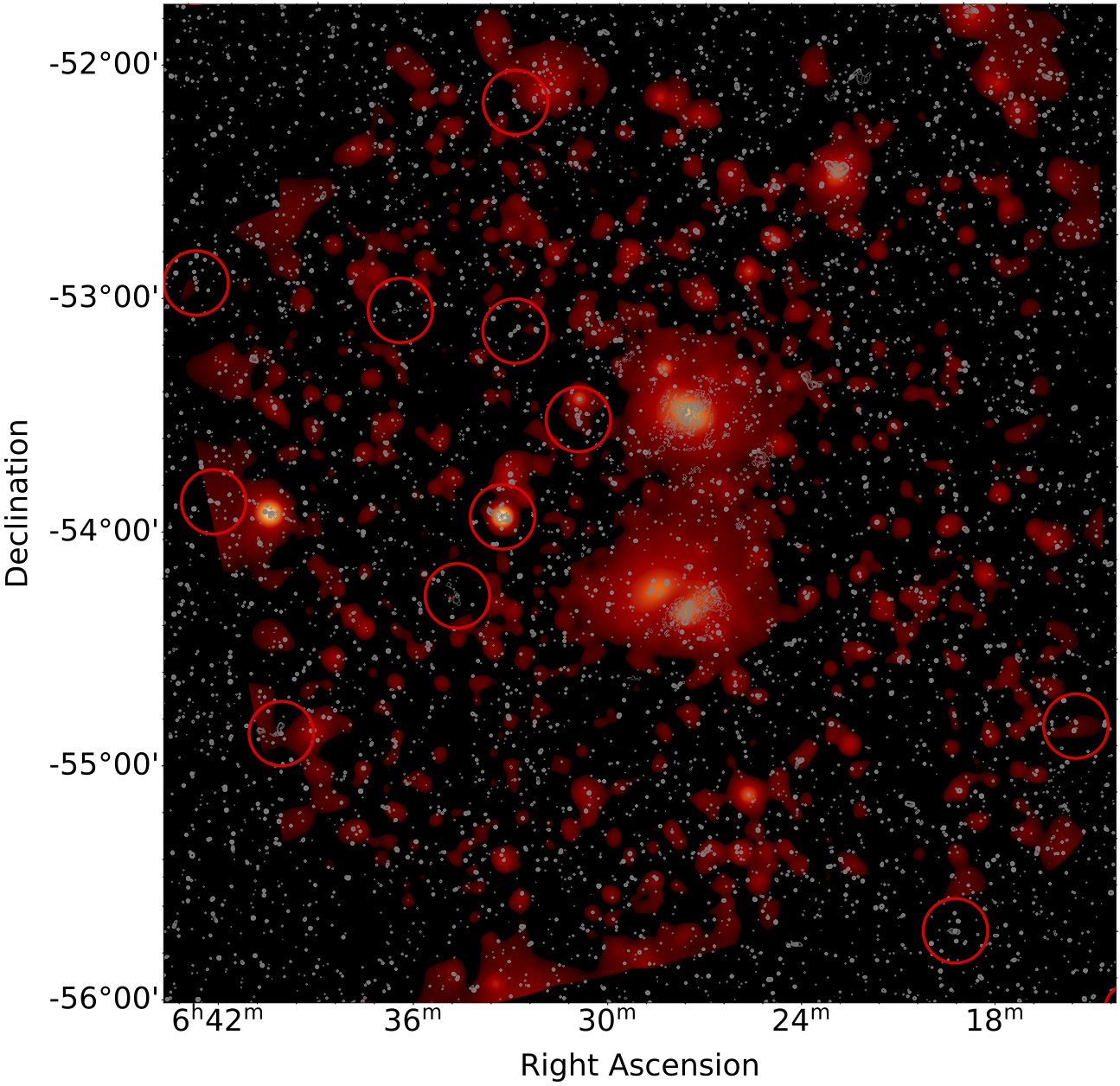}}
\caption{eROSITA X-ray 0.3--2.0~keV adaptively smoothed image of the A3391-A3395 field overlaid with the ASKAP/EMU radio contours (smoothed with a five-pixel Gaussian). The red circles show the locations of the GRGs within the FOV of the eROSITA observations. This is only the central half of the area of the ASKAP image, and it therefore only shows 11 of the 27 GRGs in 
Table 4. These GRGs in this field have higher redshifts than the A3391-A3395 cluster (see Table~\ref{tab:RGs}). Except for
perhaps two exceptions, there is no indication that these GRGs are correlated with X-ray emission peaks. The GRG with the strongest X-ray
peak is J0632-5404.}
\label{fig:large_overlay2}
\end{figure*}

Those GRGs larger than 1 Mpc that we identify in this field are listed in Table~\ref{tab:RGs}. The GRGs in the area of the ASKAP field covered also by eROSITA are marked in Fig.~\ref{fig:large_overlay2}, and radio-optical overlays for a subset of GRGs are shown in Fig.~\ref{fig:radiogal}. Photometric redshifts were taken from \cite{2016ApJS..225....5B} and from \cite{2019ApJS..242....8Z}. We measured the flux densities, $S$, of the GRGs listed in Table~\ref{tab:RGs}  within a region marking the 2$\sigma$ contour line using the local rms for $\sigma$. We used \href{https://gist.github.com/Sunmish/198ef88e1815d9ba66c0f3ef3b18f74c}{fluxtools.py} to measure the flux density and error on the flux density when considering the rms.

During our search we also identified a number of relics and cluster halos (see Table~\ref{tab:relics}). In order to identify the respective host galaxies we used optical and infrared surveys, in particular the Dark Energy Survey (DES; \cite{2020PhRvD.102b3509A}), 2MASS (\cite{2006AJ....131.1163S}), and WISE (\cite{2010AJ....140.1868W}). In the following we briefly describe some of the most remarkable objects.

\subsubsection{EMU ES J0621--5217}
\begin{figure*}
\centering
  \includegraphics[width=8cm]{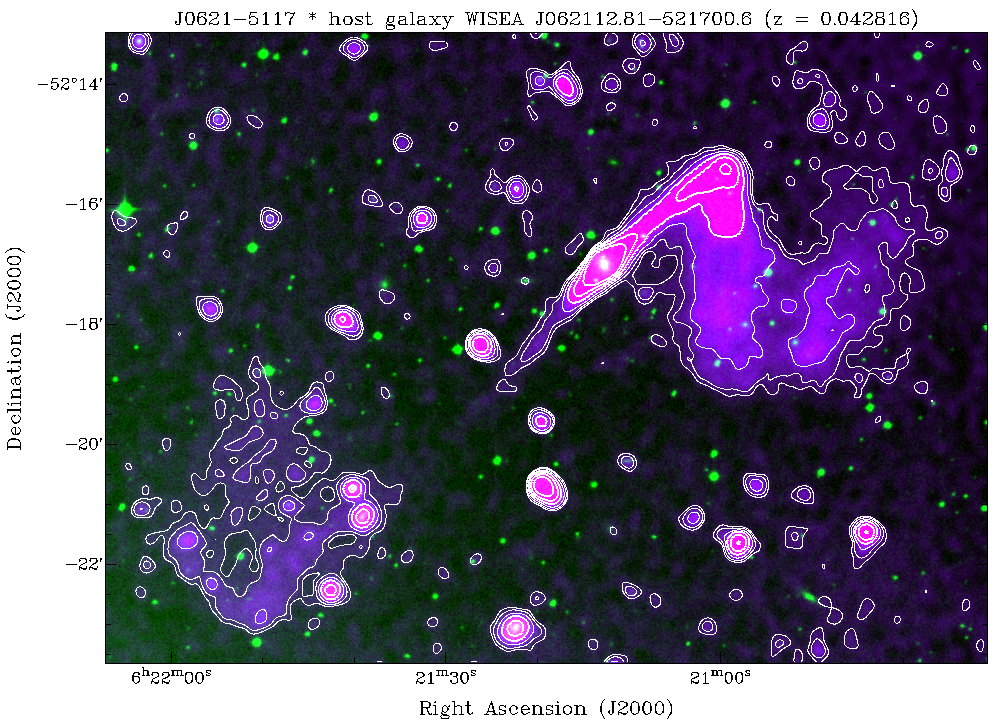}
  \includegraphics[width=8cm]{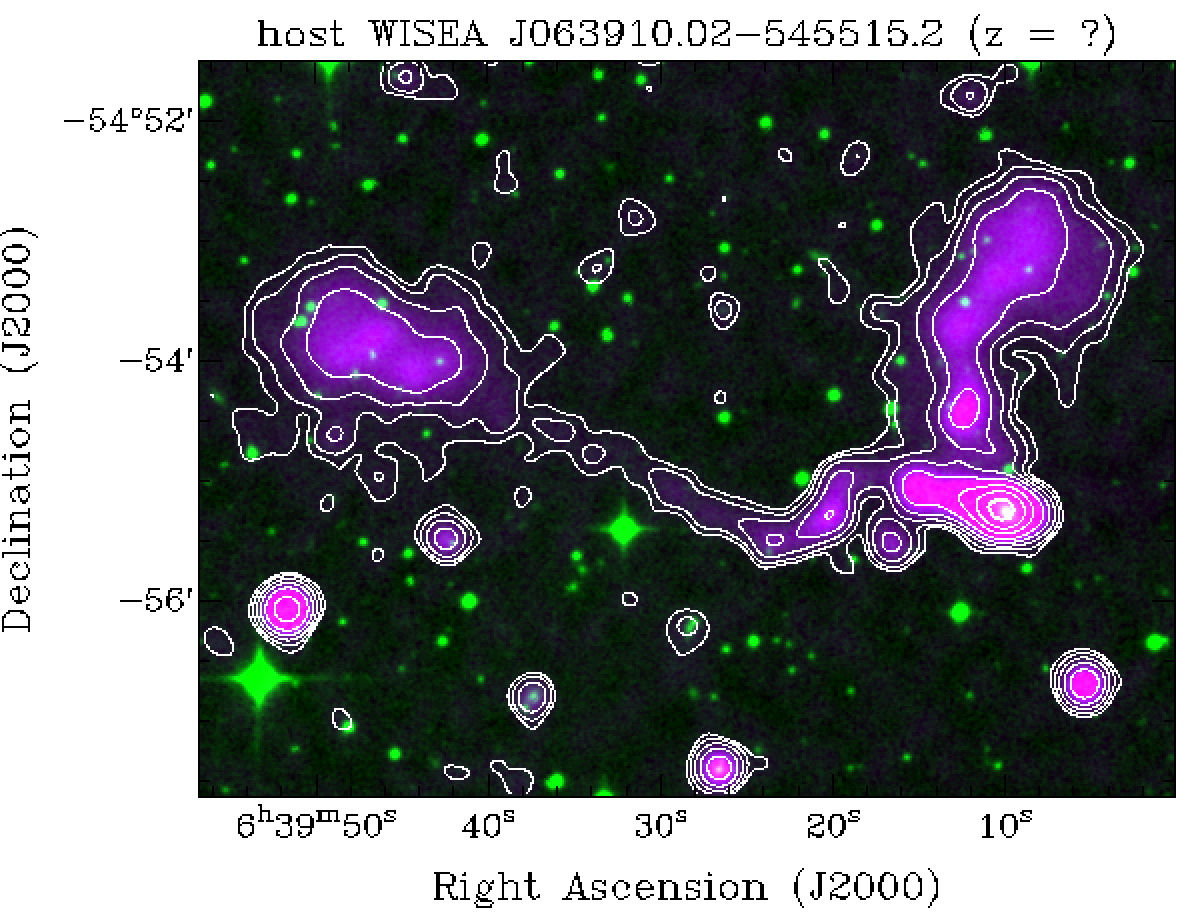}
  
  \includegraphics[width=0.8\textwidth]{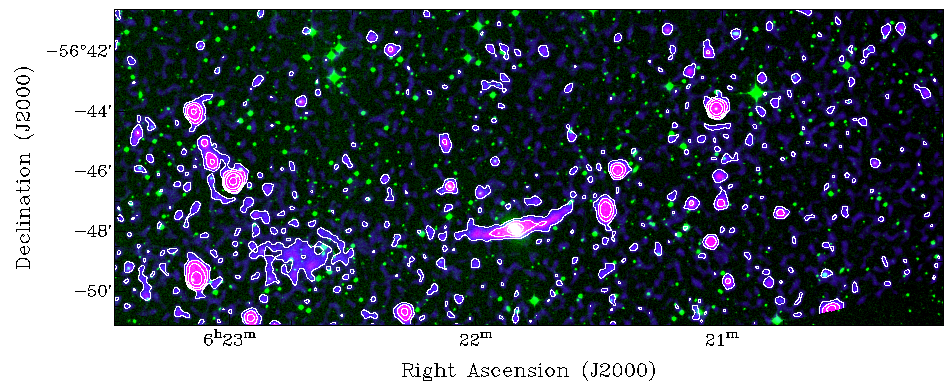}
\caption{ASKAP 1 GHz radio continuum images of the radio galaxies EMU ES J0621--5217 (top left), EMU ES J0639--5455 (top right), and EMU ES J0621--5647 (bottom) overlaid on DSS2 $R$-band images. In the top images, the radio contour levels are 0.006, 0.012, 0.025 0.05, 0.1, 0.25, 0.5, 1, and 2 mJy/beam and 0.1, 0.5, 2, 10, and 50 mJy/beam for the bottom image. The host of EMU ES J0621--5647 (bottom) is the elliptical galaxy 2MASX J06215057--5647566 ($z = 0.0539$). The very faint radio lobes appear to extend over more than $\sim15$\,arcmin.  The convolved beam is 15 arcsec.}
\label{fig:radiogal}
\end{figure*}

EMU ES J0621--5217 is an FR\,I-type radio galaxy with LAS$\sim15$\,arcmin 
with a radio core of  18.2 mJy/beam peak brightness, inner twin jets,
and evidence for precession given the shape of its outermost lobes (see Fig.~\ref{fig:radiogal}, left panel). It is highly asymmetric, likely due to projection effects, with the western lobe much brighter and closer to the core than the eastern lobe. We identify the host galaxy as WISEA J062112.81--521700.6 ($z$ = 0.042816; 6dF \cite{2009MNRAS.399..683J}). The projected linear extent of J0621--5217 is $\sim$760 kpc. This galaxy is located in the foreground of A3395-A3391.

\begin{figure*}
\centering
\includegraphics[width=8cm]{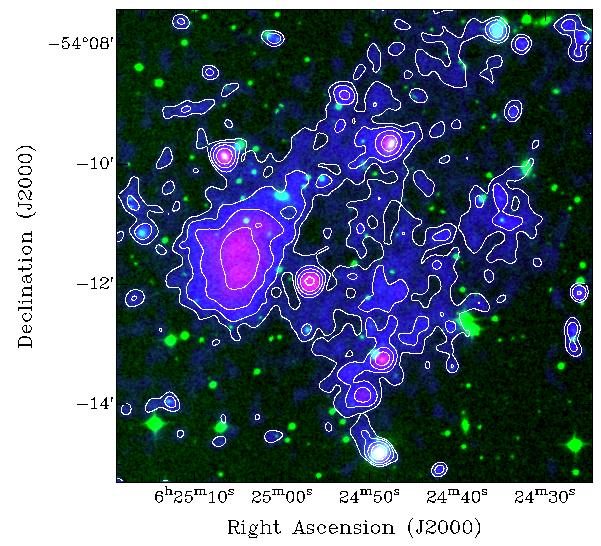} 
\includegraphics[width=8cm]{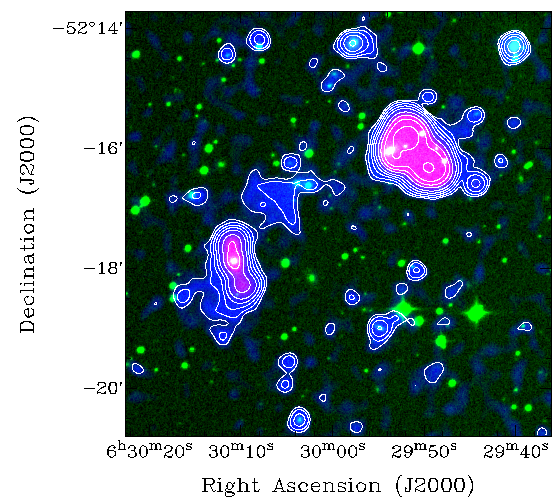} 
\includegraphics[width=8cm]{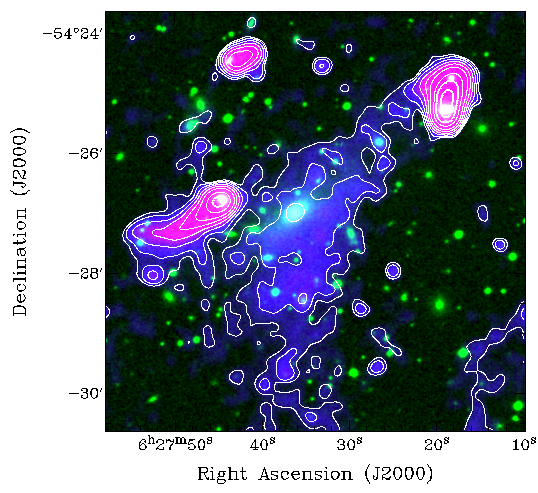}
\includegraphics[width=8cm]{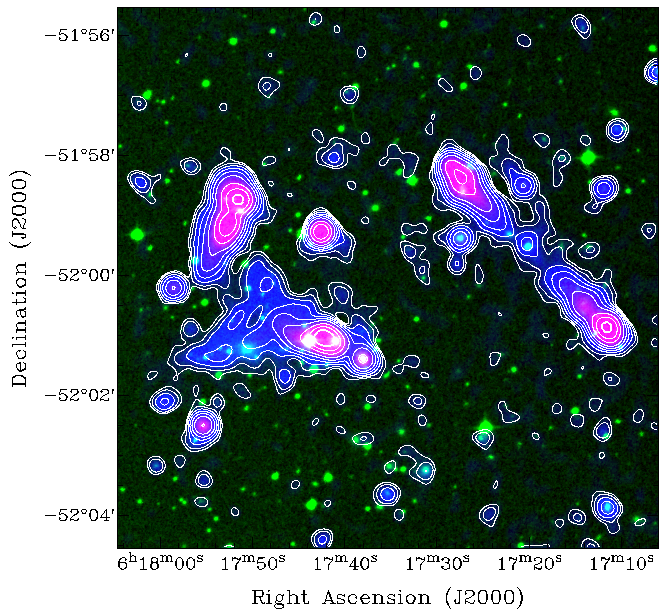}
\caption{ASKAP 1 GHz radio continuum images of some of the diffuse radio sources listed in Table~5: EMU ES J0624--5411 (A3395, top left), EMU ES J0630-5216 (A3397A/B, top right), ESO 161-G008 (in A3395N, $z$ = 0.0486, bottom left), and EMU ES J0617--5201 in A3385 (bottom right), overlaid on a DSS2 $R$-band image. In the lower right panel, the right half of the overlay shows the FRII radio galaxy DES J061722.22-515919.1 (Sect. 4.8.3). The radio contour levels are 0.006, 0.012, 0.025, 0.05, 0.1, 0.25, 0.5, 1, 2, 5, 10, and 20 mJy/beam. The convolved beam is 15 arcsec.}
\label{fig:halos}
\end{figure*}

\subsubsection{EMU ES J0644--5152}
EMU ES J0644--5152 has the largest LLS of all GRGs catalogued in this field with radio lobes extending nearly 5 arcmin and FR\,II morphology. Its host galaxy is SMSS J064437.48--515205.5, possibly a QSO, with a photometric redshift of $z \sim 0.75$ as estimated from the WISE colours (see Fig. 2 in \citealt{2018ApJS..235...10K}). We estimate its largest linear size (LLS) as 2.15 Mpc.

\subsubsection{EMU ES J0639--5455}

EMU ES J0639--5455 is a GRG with a rather unusual, WAT-like morphology (see Fig.~\ref{fig:radiogal}, right panel). Its host galaxy (2MASX J06391009--5455156, WISEA J063910.02--545515.2) is located at the radio-brightest spot in the far WSW and has a photometric redshift of $\sim$0.142,  thus well behind A3391-A3395 \citep{2016ApJS..225....5B}. 
We measure an angular extent of 7.15 arcmin, corresponding to a projected linear extent of 1.07 Mpc, a size very rarely seen for sources of this radio morphology.

\subsubsection{EMU ES J0621--5647}
EMU ES J0621--5647 is a FR\,I-type GRG, possibly extending at least $\sim$15 arcmin and possibly $\sim$32 arcmin (see Fig.~\ref{fig:radiogal}). Higher radio sensitivity is required to confirm the presence and extent of its faint radio lobes due ESE and WNW. We identify 2MASX J06215057--5647566
as the host galaxy at a redshift of $z=0.05391$ (6dF). EMU ES J0621--5647 lies at the
southern edge of the ASKAP field, but at the same redshift as A3391/3395.
If the outer lobes are confirmed, the LLS of J0621--5647 is $\sim$2\,Mpc.


\subsubsection{Summary of Giant Radio Galaxy properties}

In Table \ref{tab:RGs} we list the basic properties of the 20 GRGs (LLS$>$ 1 Mpc) and 7 candidate
GRGs found in the ASKAP field. Here we compare some of these properties
with 213 GRGs larger than 1\,Mpc as compiled from the literature
by \cite{2018ApJS..238....9K}. While the median redshift of the latter
GRGs is 0.248, our 27 GRGs have a median of 0.6, albeit with the caveat
that only two of them have spectroscopic redshifts.

We adjust the 1.4-GHz radio luminosities listed in 
\cite{2018ApJS..238....9K}
to the ASKAP observing frequency of 1.0\,GHz by assuming an average radio
spectral index
of $\alpha=-0.8$, which implies a 1-GHz flux 1.3 times larger than the
1.4-GHz
flux, or $\log P_{\rm 1GHz} = \log P_{\rm 1.4GHz}+0.12$. While the 213 GRGs from the
literature have a median $\log P_{\rm 1GHz}$[W/Hz]=25.6, our 27 GRGs have a
median
$\log P_{\rm 1GHz}$[W/Hz]=25.5, and the two distributions are statistically
indistinguishable, ranging from $\sim$23.3 to $\sim$27.4.  Consistent with
the literature sample we find no trend for the median linear size or the
median radio luminosity to change with redshift. Together with the fact
that our much smaller sample has a much larger median redshift,
we conclude that there is no evidence for cosmological evolution of the
population of GRGs.

If we exclude  four GRGs  with complex or hybrid radio morphologies from our sample of 27
and divide the rest into three classes of radio morphology, we
find median values of $\log P_{\rm 1GHz}$[W/Hz]=24.0 for the 3 FR\,Is,
24.5 for 7 FR\,IIs with remnant-type lobes,
and 26.1 for the 15 clear-cut FR\,IIs.

\subsection{Diffuse radio sources}

We discovered a number of candidate diffuse radio sources in the ASKAP field which are listed in Table~\ref{tab:relics}. The data are not sufficient for a reliable classification but we discuss the most interesting sources below. The diffuse sources in the cluster A3404 are the subject of a forthcoming paper and are not discussed here. 

\subsubsection{Abell~S0592}
\cite{2020arXiv200601833W}  detected a giant radio halo in the nearby cluster Abell S0592 ($z$ = 0.2216), also known as SPT CL J0553--3342 and MACS J0553.4--3342, using the same Early Science data  used here.
The radio halo has a diameter of about 1 Mpc and an integrated flux density of $S_{\rm 1013 MHz} = 9.95 \pm 2.16$ mJy. The diffuse emission is also seen in ATCA data at 2.215 GHz.

 \subsubsection{Abell~3395E}
We find extended radio emission associated with the bright elliptical galaxy ESO 161-G008 ($z$ = 0.0486), the central galaxy of A3395E. This could be diffuse emission from a radio halo, although image artefacts in that area make the full size of the structure difficult to assess. Three other cluster ellipticals in the area also have radio emission with significant tails: WISEA J062744.63--542644.4 (PGC~019090; $z$ = 0.044281) has a tail of 2 arcmin in length  towards the east, while DES~J062719.2--542515.24 ($z$ = 0.0456) has a tail of 1 arcmin pointing north and 2MASS J06274388--5424266 ($z$ = 0.044971) has a short tail in the NW direction. The aforementioned galaxies lie within the X-ray cocoon of the cluster, which is aligned with and centred on ESO 161-G008. There are at least 40 known cluster
members within 5 arcmin of ESO 161-G008.

\subsubsection{Abell~3385}
We find diffuse emission in Abell~3385 ($z$ = 0.1245; 6df) also known as 1RXS J061747.8--520132. This emission lies right at the NW edge of the eROSITA FOV. North of the 1RXS position is a bright radio source that appears to connect to two faint strands of radio emission that run from the NW to the SE. The angular length of the filaments is 2 arcmin which corresponds to 270 kpc. The flux density of this diffuse source is $S_{\rm 1013 MHz} = 13.7 \pm 2.3$ mJy. This source is likely to be a radio relic. Finally, on the right (W) half of the bottom-right panel in Fig.~\ref{fig:halos} we see
the regular and straight FR II source hosted by DES J061722.22-515919.1,
a likely member of A3385, with $z_{\rm ph}\sim 0.128$, LAS$\sim 4.0'$, and LLS$\sim 550$ kpc.

\subsubsection{Abell~3397}
A3397 is a poorly studied galaxy cluster that is located north of A3391-A3395 with no published redshift, but the 6dF redshift survey \citep{2009MNRAS.399..683J} shows seven galaxies with mean measured redshift $\langle z\rangle =0.715$ (A3397A) and four others with $\langle z \rangle=0.1063$ (A3397A) within one Abell radius (1.7$'/z$) from the Abell centre \citep{2014MNRAS.445.4073C}. One more galaxy, 2MASX J062940.23-521418.5, with a redshift measured by \cite{2014ApJ...797...82L}, makes a total of eight spectroscopic members of A3397A 
  with a velocity dispersion of $750 \pm 80$ km\,s$^{-1}$ based on the BIweight
scale of ROBUST \citep{1990AJ....100...32B}. The eROSITA image shows an extension towards the west that is indicative of a merger in the east--west direction. The X-ray image also shows an extension towards the north. An extended radio source lies on the western side of the X-ray peak with hints of diffuse emission (see Fig.
~\ref{fig:halos}, top right panel) and appears to be composed of several components; its LAS is 1.4 arcmin which corresponds to an LLS of 600 kpc at $z=0.715$. The flux density of this source is $S_{\rm 1013 MHz} = 0.23 \pm 0.04$ Jy. The nature of the source is unclear.

\subsubsection{EMU ED J0624-5414}

The source sits in a patch of diffuse radio emission whose angular extent is about 4 arcmin. The redshift of WISEA J0624-5414 is $z=0.045$. It appears as two connected bars of emission that stretch from the SE to the NW. The main blob in the northern part of the structure extends for about 1.8 arcmin before it fades quite abruptly. The flux density of this source is $S_{\rm 1013 MHz} = 0.13 \pm 0.019$ Jy and there is no counterpart in the eROSITA image. This source is likely to be a radio halo (see Fig.~\ref{fig:halos}, upper left panel). 

\begin{table*}
\centering
\caption{Properties of GRGs (LLS $>$ 1 Mpc) discovered in the 30 deg$^2$ ASKAP field towards Abell 3391/5 sorted in decreasing order of LLS. We list the host galaxy name, their spectroscopic (s), photometric (p), or estimated (e) redshifts, flux densities, $S$, luminosities, $\log P$(1013~MHz)/[W/Hz], and radio morphology. The superscript `C' after the source name denotes a GRG candidate for which we list the most likely host name. Flux densities, $S$, were derived by integration of the ASKAP-DD image and measured within the $3\sigma$ contour level where $\sigma$ is the local rms. Flux densities, $S$, marked with $^{\dag}$ were measured at $2\sigma$.}
\begin{tabular}{lllccccccc}
\hline
 EMU ES  & Host name & Redshift & \multicolumn{2}{c}{Extent of radio lobes} & $S$(1013~MHz) & $\log P$ & Radio\\
 Name  & $\alpha,\delta$(J2000) & $z$~~~~type & LAS & LLS & mJy & & morphology& \\
 &  & & [$'$] & [Mpc] & \\
\hline
J0644--5152 & SMSS J064437.48--515205.5 & 0.75~~~~p & 4.89 &  2.15 & $17.6 \pm 1.8$& 25.6 & FR\,II \\
J0621--5647 & 2MASX J06215057--5647566  & 0.0539s & 32? & 2.0? & $23.3 \pm 5.8$ $^{\dag}$& 23.3 & FR\,I/II remn., WAT? \\
J0608--5409 $^\text{C}$ & DES--J060805.04--540918.2  & 1.2~~~~~~p &  3.7 &  1.84 & $21.5 \pm 2.2$ & 26.2 & FR\,II  \\ 
J0634--5309 & DES~J063428.15--530928.4  & 0.34~~~~p &  $>$5.6 &  $>$1.6 & $4.5 \pm 0.5 $ $^{\dag}$& 24.2 & FR\,II remn.  \\
J0612--5157 $^\text{C}$ & DES--J061215.47--515738.1 & 0.7~~~~~~e & 3.7 & 1.59 & $54.4 \pm 5.4$& 26.0 & FR\,I/II   \\
\\
J0613--5117 $^\text{C}$ & DES--J061339.02-511703.5  & 1.8~~~~~~e & 3.1 & 1.57 & $5.1 \pm 0.5$& 25.9 & FR\,II asym. \\
J0639--5136 & VHS J063919.22--513652.1 & 0.45~~~~e & 4.16 & 1.44 & $41.2 \pm 4.1$&25.5 & FR\,II \\
J0640--5257 & WISEA J064013.78--525702.2 & 0.6~~~~~~e & 3.57 & 1.43 & $9.9 \pm 1.0$& 25.1& FR\,II relic \\
J0633--5424 & 2MASX J06333235--5424190 & 0.165~~p & 7.65 & 1.30 & $20.8 \pm 2.1$& 24.2& FR\,I  \\
J0613--5621 & DES~J061337.62--562114.6 & 0.35~~~~p & 4.3 & 1.27 & $7.9 \pm 0.8$ $^{\dag}$& 24.5 & FR\,I/II remn. \\
\\
J0629--5341 & DES~J062935.40--534124.4 & 1.65~~~~p & 2.48 & 1.26 & $ 102.0 \pm 10.3 $& 27.2 & FR\,II \\
J0636--5616 & DES~J063633.92--561635.6 & 0.6~~~~~~e & 3.06 & 1.23 & $10.8 \pm 1.1$& 25.2 & FR\,II \\
J0641--5136 & SMSS J064130.39--513602.7 & 0.395~~p & 3.67 & 1.18 & $6.1 \pm 0.6$& 24.5 & FR\,I/II \\
J0619--5558 & DES~J061909.04--555843.7 & 1.3~~~~~~p & 2.33 & 1.17 & $98.6 \pm 9.9$& 26.9 & FR\,II cpx \\
J0630--5218 & DES~J063043.12--521821.2  & 1.8~~~~~~p & 2.27 & 1.15 & $19.6 \pm 2.0$& 26.5 & FR\,II \\
\\
J0610--5412 $^\text{C}$ & DES--J061042.30--541225.8 & 0.8~~~~~~e & 2.5? & 1.13? & $1.2 \pm 0.1 $ $^{\dag}$& 24.5 & FR\,II?  \\
J0631--5317 & DES~J063113.59--531719.6 & 0.45~~~~e & 3.27 & 1.13 & $27.5 \pm 2.8$& 25.3 & FR\,II \\
J0615--5506 $^\text{C}$ & DES--J061518.25-550647.5 & 1.015~~p & 2.33 & 1.13 & $51.2 \pm 5.1$ & 26.4 & FR\,II, no core? \\
J0632--5404 & 6dFGS gJ063201.2--540458 & 0.193~~s & 5.80 & 1.12 & $556.0 \pm 55.9$& 25.8 & FR\,II plume \\
J0649--5501 & SMSS J064947.02--550109.6 & 0.20~~~~p & 5.49 & 1.09 & $370.6 \pm 37.2$& 25.6 & FR\,II, X-shaped \\
\\
J0640--5353 & WISEA J064024.95--535347.1 & 0.8~~~~~~~e & 2.38 & 1.07 & $81.6 \pm 8.2$& 26.3 & FR\,II \\
J0621--5638 & DES~J062133.55--563822.9 & 0.321~~~e & 3.7? & 1.03 & $ 5.4 \pm 0.5$ $^{\dag}$& 24.2& FR\,II remn. \\
J0625--5137 $^\text{C}$ & DES--J062522.66--513750.1 & 0.6~~~~~~~e & 2.5 & 1.00 & $2.3 \pm 0.2$ & 24.5 & FR\,II remn. \\
J0639--5455 & 2MASX J06391009--5455156 & 0.142~~~p & 7.15 & 1.07 & $61.2 \pm 6.1$& 24.5 & FR\,I\\
J0615--5136 & 2MASX J06153812--5136193 & 0.126~~~p & 7.9 & 1.07 & $43.8 \pm 4.4$& 24.3 & FR\,II relic \\
\\
J0634--5644 & WISEA J063429.02-564436.0 & 1.2~~~~~~~e & 2.06 & 1.02 & $41.2 \pm 4.1$ & 26.4 & FR\,II \\
J0636--5646 $^\text{C}$ & DES J063630.11-564617.7 & 0.832~~~p & 2.32 & 1.06 & $95.6 \pm 9.6$ & 26.4 & FR\,II \\
\hline 
\end{tabular}
\label{tab:RGs}
\end{table*}

\begin{table*}
\centering
\caption{Properties of diffuse radio sources discovered in the 30 deg$^2$ ASKAP field towards A3391/95.  We list the spectroscopic (s) or photometric (p) redshifts. A question mark in the second column indicates uncertainty as to whether the cited object is the host of the radio source.}
\begin{tabular}{cccccccccc}
\hline
 EMU ES& centre position & \multicolumn{2}{c}{redshift} & \multicolumn{2}{c}{extent} \\
 Name &  & $z$ & type & [arcmin] & [kpc] & type & comments \\
\hline
J0617--5201 & WISEA J061750.63--520113.8 & 0.1245 & s & 4? &  550 & cluster relics & A3385 \\
J0624--5411 & WISEA J062447.61--540939.6 ? & 0.0490 & s & 2.5? & 150? & relic or halo ? & A3395 substructure \\
J0627--5426 & ESO 161-G008 & 0.0486 & s & 3? & 170? & radio halo? & A3395E \\
J0628--5448 & ... & ? & & 3+ & 200? & relic & in A3395 ? \\
J0630--5216 & WISEA J063003.92--521634.0 & 0.0765 & p & 4?  & ? &  ? & A3397A ? \\
~~~~~" & WISEA J063002.67--521636.3 & 0.0874 & p & "  & ? &  ? & A3397B ? \\
J0638--5358 & PSZ2 G263.14-23.41& 0.2220 & s & 3 & 650 & radio halo & AS0592 \\
J0645--5413 & 2MASX J06452948--5413365 & 0.1670 & s & 5? & 850 & radio halo + NATs & A3404 \\
\hline 
\end{tabular}
\label{tab:relics}
\end{table*}

\section{Conclusions}
\label{sec:conclusions}

We present 1 GHz radio observations of the pre-merging system of galaxy clusters A3391-A3395 with ASKAP/EMU. Accompanying eROSITA observations of this system yield the best large-scale X-ray characterisation of the emission bridge between A3391 and A3395 to date (Reiprich et al., submitted).

We do not detect diffuse radio emission in the X-ray bridge between these two clusters. Provided that the acceleration mechanism is Fermi-II acceleration, the non-detection implies that the turbulent velocity in this system is smaller than in the A399-401 system, possibly because of the smaller mass.  This is something that can be verified with the future ATHENA X-ray mission \citep{2013arXiv1306.2307N}.

Moreover, we find a plethora of other interesting radio sources in a 30 deg$^2$ field around A3391-A3395. We identified around 200 objects extending over more than 1 arcmin. Of these, we present a list of 27 that are GRGs 
with linear projected sizes of greater than 1 Mpc (see Table~\ref{tab:RGs}). This surface density is four times that found by \cite{2020A&A...635A...5D} who find about 0.2 GRGs (LLS $>1$ Mpc per deg$^2$ based on LoTSS DR1. For galaxies with LLS $>0.7$ Mpc, we obtain a sky density of $\approx 1.7$ deg$^{-2}$, three times higher than previously reported \citep{2020A&A...635A...5D}. The 1.4-GHz luminosities of our sample of GRGs are consistent with distributions found in the literature. Our sample shows no evidence in support of cosmological evolution of the GRG population.
  
Finally, we identify seven candidates for radio relics and halos, which we list in Table~\ref{tab:relics} together with previously published examples \citep{2020arXiv200601833W}. It will be important to follow these up with deeper observations and to make spectral index maps in order to classify the sources and understand their origin. This will be the subject of future work.

\begin{acknowledgement}
This work is based on data from eROSITA, the primary instrument aboard SRG, a joint Russian-German science mission supported by the Russian Space Agency (Roskosmos), in the interests of the Russian Academy of Sciences represented by its Space Research Institute (IKI), and the Deutsches Zentrum f\"ur Luft- und Raumfahrt (DLR). The SRG spacecraft was built by Lavochkin Association (NPOL) and its subcontractors, and is operated by NPOL with support from the Max Planck Institute for Extraterrestrial Physics (MPE).

The development and construction of the eROSITA X-ray instrument was led by MPE, with contributions from the Dr. Karl Remeis Observatory Bamberg \& ECAP (FAU Erlangen-N\"urnberg), the University of Hamburg Observatory, the Leibniz Institute for Astrophysics Potsdam (AIP), and the Institute for Astronomy and Astrophysics of the University of T\"ubingen, with the support of DLR and the Max Planck Society. The Argelander Institute for Astronomy of the University of Bonn and the Ludwig Maximilians Universit\"at Munich also participated in the science preparation for eROSITA.

The eROSITA data shown here were processed using the eSASS software system developed by the German eROSITA consortium.
The Australian SKA Pathfinder is part of the Australia Telescope National Facility which is managed by CSIRO. Operation of ASKAP is funded by the Australian Government with support from the National Collaborative Research Infrastructure Strategy. ASKAP uses the resources of the Pawsey Supercomputing Centre. Establishment of ASKAP, the Murchison Radio-astronomy Observatory and the Pawsey Supercomputing Centre are initiatives of the Australian Government, with support from the Government of Western Australia and the Science and Industry Endowment Fund. We acknowledge the Wajarri Yamatji people as the traditional owners of the Observatory site. The Australia Telescope Compact Array (/ Parkes radio telescope / Mopra radio telescope / Long Baseline Array) is part of the Australia Telescope National Facility which is funded by the Australian Government for operation as a National Facility managed by CSIRO. This paper includes archived data obtained through the Australia Telescope Online Archive (http://atoa.atnf.csiro.au). This work was supported by resources provided by the Pawsey Supercomputing Centre with funding from the Australian Government and the Government of Western Australia. We acknowledge and thank the builders of ASKAPsoft.
Support for the operation of the MWA is provided by the Australian Government (NCRIS), under a contract to Curtin University administered by Astronomy Australia Limited. 
MB acknowledges support from the Deutsche Forschungsgemeinschaft under Germany's Excellence Strategy - EXC 2121 "Quantum Universe" - 390833306. 
HA benefited from grant CIIC 90/2020 of Universidad de Guanajuato, Mexico. AB and DNH acknowledge support from the ERC StG DRANOEL 714245. AB acknowledges support from the MIUR grant FARE "SMS". SWD acknowledges an Australian Government Research Training Program scholarship administered through Curtin University. LR receives support from the U.S. National Science Foundation grant AST17-14205 to the University of Minnesota. The authors made use of the database CATS \citep{2005BSAO...58..118V} of the Special Astrophysical Observatory.
\end{acknowledgement}


\bibliographystyle{aa}
\bibliography{example}

\end{document}